\documentclass[12pt]{article}
\usepackage{graphicx}  % needed for figures
\usepackage{subfigure}
\usepackage{dcolumn}   % needed for some tables
\usepackage{bm}        % for math
\usepackage{amssymb}   % for math
\usepackage{ulem}
\usepackage{dcolumn}
\usepackage{epsfig}
\usepackage{array}
\usepackage{url}
\newcommand{\gev}  {\mbox{${\rm GeV}$}}

\newcommand{\invfb}{\mbox{${\rm fb}^{-1}$}}

\newcommand{\Br}  {\mbox{${\cal B}$}}   % Branching ratio

\newcommand{\etal} {\mbox{$et~al.$}}

\newcommand{\pt}  {\mbox{$p_{\rm T}$}}
\newcommand{\ptvis}  {\mbox{$p_{\rm T}^{\rm vis}$}}
\newcommand{\et}  {\mbox{$E_{\rm T}$}}

\newcommand{\met} {\mbox{${E\!\!\!\!/_{\rm T}}$}}

\newcommand{\meff}{\mbox{$M_{{\rm eff}}$}}
\newcommand{\meffpeak}{\mbox{$M_{{\rm eff}}^{\rm peak}$}}
\newcommand{\meffb}{\mbox{$M_{{\rm eff}}^{(b)}$}}
\newcommand{\meffbpeak}{\mbox{$M_{{\rm eff}}^{(b)\; \rm peak}$}}
\newcommand{\mefftwob}{\mbox{$M_{{\rm eff}}^{(2b)}$}}
\newcommand{\mefftwobpeak}{\mbox{$M_{{\rm eff}}^{(2b)\; \rm peak}$}}

\newcommand{\bbar} {\mbox{$\overline{b}$}}
\newcommand{\tbar} {\mbox{$\overline{t}$}}

\newcommand{\bbbar}{\mbox{$b\overline{b}$}}

\newcommand{\azero}{\mbox{$A_{0}$}}
\newcommand{\tanb} {\mbox{$\tan\beta$}}
\newcommand{\mzero}{\mbox{$m_{0}$}}
\newcommand{\mhalf}{\mbox{$m_{1/2}$}}

\newcommand{ \gluino}   {\mbox{$\tilde{g}$}}
\newcommand{ \squark}   {\mbox{$\tilde{q}$}}
\newcommand{ \squarkL}   {\mbox{$\tilde{q}_{L}$}}
\newcommand{ \squarkR}   {\mbox{$\tilde{q}_{R}$}}

\newcommand{ \usquarkL}  {\mbox{$\tilde{u}_{L}$}}
\newcommand{ \usquarkR}  {\mbox{$\tilde{u}_{R}$}}

\newcommand{ \sbottomone}{\mbox{$\tilde{b}_{1}$}}

\newcommand{ \sbottomtwo}{\mbox{$\tilde{b}_{2}$}}

\newcommand{ \stopone}  {\mbox{$\tilde{t}_{1}$}}

\newcommand{ \stoptwo}  {\mbox{$\tilde{t}_{2}$}}

\newcommand{ \seleR}    {\mbox{$\tilde{e}_{R}$}}

\newcommand{ \seleL}    {\mbox{$\tilde{e}_{L}$}}

\newcommand{ \stauone}  {\mbox{$\tilde{\tau}_{1}$}}

\newcommand{ \stautwo}  {\mbox{$\tilde{\tau}_{2}$}}
\newcommand{ \stau}     {\mbox{$\tilde{\tau}$}}

\newcommand{ \schionezero }{\mbox{$\tilde{\chi}_{1}^{0}$}}
\newcommand{ \schitwozero }{\mbox{$\tilde{\chi}_{2}^{0}$}}

\newcommand{ \schionepm }{\mbox{$\tilde{\chi}_{1}^{\pm}$}}

\newcommand{ \pythia } {{\tt PYTHIA}}
\newcommand{ \isasugra } {{\tt ISASUGRA}}
\newcommand{ \isajet }    {{\tt ISAJET}}
\newcommand{ \pgs }    {{\tt PGS4}}

\newcommand{ \DMrelic }{\mbox{$\Omega_{\schionezero}h^{2}$}}
\newcommand{ \pncross }{\mbox{$\sigma_{p-\schionezero}$}}

\makeatletter
\def\alt{\mathrel{\mathpalette\gl@align<}}
\def\agt{\mathrel{\mathpalette\gl@align>}}
\def\gl@align#1#2{\lower.6ex\vbox{\baselineskip\z@skip\lineskip\z@
\ialign{$\m@th#1\hfil##\hfil$\crcr#2\crcr\sim\crcr}}}
\makeatother

\begin{document}
\begin{flushright}
%{\tt }\\
MIFP-08-20 \\
ACT-04-08

\end{flushright}
\vspace*{2cm}
\begin{center}
{\baselineskip 25pt \large{\bf
Supersymmetry Signals of Supercritical String Cosmology at the Large Hadron Collider
} \\

}

\vspace{1cm}

{\large
Bhaskar~Dutta$^{1}$, Alfredo~Gurrola$^{1}$, Teruki~Kamon$^{1}$, Abram~Krislock$^{1}$, A.B.~Lahanas$^{2}$, N.E.~Mavromatos$^{3}$, D.V.~Nanopoulos$^{1,4,5}$
} \vspace{.5cm}

{
\it $^{1}$Department of Physics, Texas A\&M University, College
Station,
TX 77843-4242, USA\\
$^{2}$ University of Athens, Physics Department, Nuclear and Particle
Physics Section, GR-157 71, Athens, Greece\\
$^{3}$ King's College London, University of London, Department of Physics,
Strand WC2R 2LS, London, U.K.\\
$^{4}$Astroparticle Physics Group, Houston Advanced Research Center
(HARC),  Mitchell Campus, Woodlands, TX 77381, USA\\
$^{5}$Academy of Athens, Division of Natural Sciences,
28 Panepistimiou Avenue, Athens 10679, Greece
}
\vspace{.5cm}

\vspace{1.5cm} {\bf Abstract}\end{center}

We investigate the minimal supergravity (mSUGRA) signals at the LHC in the context of supercritical string cosmology (SSC).  In this theory, the presence of a time dependent dilaton  provides us with a smoothly evolving dark energy and modifies the dark matter allowed region of the mSUGRA model with standard cosmology.  Such a dilaton dilutes  the supersymmetric dark matter density (of neutralinos) by a factor $\mathcal{O}(10)$ and consequently the regions with too much dark matter in the standard scenario are allowed in the SSC. The final states expected at the LHC in this scenario, unlike the standard scenario, consist of $Z$ bosons, Higgs bosons, and/or high energy taus. We show how to characterize these final states and determine the model parameters. Using these parameters, we determine the dark matter content and the neutralino-proton cross section. All these techniques can also be applied to determine model parameters in SSC models with different SUSY breaking scenario
 s.

\thispagestyle{empty}

\bigskip
\newpage

\addtocounter{page}{-1}

\section{Introduction}
The recent WMAP data~\cite{WMAP} has determined the content of the universe very precisely.  The dark matter and dark energy compose about 23\% and 73\% of the total energy density of the universe, respectively.

The origin of dark matter can be explained in supersymmetry (SUSY) models, where the lightest SUSY particle, the neutralino (in most SUSY models)~\cite{neuDM}, is the dark matter candidate.  SUSY combined with supergravity grand unification (SUGRA GUT)~\cite{sugra1},
resolves a number of the problems inherent in the standard model (SM).  The
SUGRA GUT model not only solves the gauge hierarchy problem and
predicts grand unification at the GUT scale $M_{\rm G} \sim 10^{16}$ GeV but also allows for the spontaneous breaking of
SUGRA at the $M_{\rm G}$ scale in a hidden sector,
leading to an array of soft breaking masses.
The renormalization group equations (RGEs) then show
that this breaking of SUGRA leads naturally to the breaking of
SU(2)$\times$U(1) of the SM at the electroweak scale~\cite{EW}.  SUSY breaking masses around a TeV for most of the SUSY parameter space are allowed by other experimental constraints. It is also very interesting to note that achieving the WMAP relic density requires the annihilation cross section of the lightest neutralino in these SUSY models to be of order 1 pb with $M_{\schionezero} \sim \mathcal{O} (100~\gev)$.  Such a mass scale is reachable at the LHC.

The origin of dark energy is not well understood.  The simplest proposal is to add a cosmological constant in Einstein's equation. However, the reason why the dark matter content is comparable to the dark energy content at the present time remains a puzzle. Another proposal is that a
quintessence scalar field is responsible for dark energy~\cite{de}. However, this requires the field to have a very small mass and is not well motivated in particle physics models. In the context of string theory, the dilaton can play the role of dark energy~\cite{ssc, destring}. One also finds proposals which involve, for example, modifications to General Relativity, Braneworld scenarios, or
Topological defects, which are invoked to explain this fundamental issue.

In this paper, we will investigate experimental signatures of SUSY as consequences of a rolling dilaton in the $Q$-cosmology scenario~\cite{ssc} which offers an alternative framework that establishes the Supercritical (or non-critical) String Cosmology (or SSC).  In the SSC framework, the dark energy has two components:  One component arises from the dilaton, $\phi$, and the other arises from the $Q^2$ which is associated with the central charge deficit. Both $Q$ and the dilaton have time dependent pieces. It was shown that the SSC scenario~\cite{ssc1} is consistent with the smoothly evolving dark
energy at least for the last ten billion years ($0 < z < 1.6$), in accordance with the very recent observations on supernovae~\cite{reiss}.

The presence of this time dependent dilaton affects the relic density calculation since it modifies the Boltzman equation in the following way:
${dn\over dt}+3 H n+\langle\sigma v\rangle(n^2-n_eq^2)-\dot\phi n=0$
The relic density is given by
\begin{equation}
\Omega h^2=R\times (\Omega h^2)_0
\label{eqOmegamod}
\end{equation}
where $R \sim \mathrm{exp}[\int^{x_f}_{x_0}(\dot\phi H^{-1}/x )dx]$ and $(\Omega h^2)_0$ denotes the relic density that is obtained by ordinary cosmology. It is possible to determine $R$ by solving for $\phi$ from the field equations for this SSC scenario.  The value of $R$ is about 0.1 in order to satisfy the recent observation of the evolution of dark energy in the range $0 < z < 1.6$. This new factor changes the  profile of dark matter allowed region in SUSY models.

To investigate the SUSY signatures we use the minimal SUGRA (mSUGRA) model and calculate the dark matter content in the context of the SSC framework. We note that the low energy limit of string theory is certainly much more complicated than mSUGRA, and there are many different effective theories, depending on the details of compactification and SUSY breaking~\cite{hetero}. The relevant dark matter phenomenological analyses are highly model dependent~\cite{mavromatoslhc}. In some cases, such as the orbifold-compactified heterotic models~\cite{hetero}, there might be situations in which the couplings of matter with stabilized dilatons lead to non-thermal dark matter, thus leading to completely different phenomenology.

However, the SSC framework is characterized by a non-stabilized dilaton which runs in cosmic time~\cite{ssc1}.  In this context, it is possible to have thermalization of weakly interacting dark matter, such as the mSUGRA lightest neutralino ($\schionezero$) which couples to the dilaton.  In this sense, the mSUGRA framework provides a sufficiently non-trivial and generic pilot study of the novel effects the running dilaton has on the abundance of thermal dark matter relics.

The mSUGRA parameters are the universal scalar mass, $\mzero$, the universal gaugino mass, $\mhalf$, the universal soft breaking trilinear coupling constant, $\azero$, the ratio of Higgs expectation values, $\tanb$, and the sign of $\mu$, the bilinear Higgs coupling constant.
In the case of the standard cosmology,
%  has four distinct regions~\cite{darkrv}: (i)~the stau neutralino
%($\stauone$-$\schionezero$) coannihilation region where
%$\schionezero$ is the lightest SUSY particle (LSP) (In Fig.~\ref{figParameterSpace}, this dark %matter allowed region is the narrow corridor along $\mhalf$ for smaller values of $\mzero$), %(ii)~the
%$\schionezero$ having a larger Higgsino component (focus point) (in Fig.~\ref{figParameterSpace}, %this dark matter allowed region appears for larger values of $\mzero$), (iii)~the scalar Higgs %($A^0$, $H^0$) annihilation funnel
%(2$M_{\schionezero}\simeq M_{A^0,H^0}$),
%(iv) a bulk
%region where none of these above properties is observed, but this
%region is now very small due to the existence of other experimental
%bounds.
if we concentrate on smaller values of $\mzero$ and $\mhalf$, then the stau neutralino ($\stauone$-$\schionezero$) coannihilation region is the only dark matter allowed region (which is also allowed by the $g_{\mu} - 2$ constraint)~\cite{darkrv}. However, due to the presence of the extra factor $R$, the parameter space in the SSC scenario requires larger values of $\mzero$.  This is because a smaller annihilation cross section is required in the presence of the dilaton contribution. The magnitude of $\mzero$ is however, still much smaller than the focus point region~\cite{focus} (in this region, the magnitude of $\mu$ is small and therefore the lightest neutralino has a large Higgsino component).  This difference in $\mzero$ will produce new types of signals at the LHC for the SSC model. For example, in the case of the standard cosmology, the allowed region for low $\mzero$ requires the $\stauone$ and $\schionezero$ to have nearly degenerate masses within $\sim10~\gev$.  This
  produces low energy $\tau$'s in the final states~\cite{LHCrelicdensity,LHCtwotau,LHCthreetau}.  In contrast, in the SSC model, $Z$
 bosons, Higgs bosons or high energy taus appear in the final states. These final states, which we will discuss for this SSC scenario, actually exist in most regions of the SUSY parameter space. Therefore,  even without any cosmological motivation, searching for these signals is a worthwhile exercise. Furthermore, even though we have used the SSC as our motivation to probe the signals of the SUGRA model, one can come up with any other cosmological framework where the Boltzmann equation is modified in such a way that the universe is not really overclosed in this wide region of SUGRA paramater space. This analysis is valid for all these scenarios. One can also use other SUSY breaking scenarios in the context of SSC. Our analysis of signals can still be applicable in those new scenarios.  However, additional observables may be required in order to determine the model parameters.

The determination of the factor $R$ in Eq.~\ref{eqOmegamod} is important since it will tell us whether we satisfy the cosmological observation for the evolution of dark energy for $0 < z < 1.6$. In order to calculate $R$ we need to calculate the relic density precisely at the collider. In this paper, we first show how to analyze and develop appropriate cuts to extract the signals in the newly allowed parameter space in order to determine model parameters.  We also construct new observables necessary for the determination of such parameters. Then, using these parameters, we determine the accuracy of the result for the dark matter content.  Finally, when the LHC will be operating, the dark matter direct detection experiments also will be probing the SUSY parameter space. The neutralino-proton scattering cross section is different for this newly allowed parameter space compared to the standard cosmology case. Thus we also determine the accuracy of the result for the neutralino p
 roton scattering cross section based on the LHC measurements.   This will be very useful when we will combine the data from these direct detection experiments with the LHC data to extract the final model.

%Although we use the mSUGRA model to show the dilution effect of the overdense region of the %standard cosmology in the SSC scenario, one can use other SUSY breaking scenarios in the context %of SSC. Our analysis of signals can still be applicable in those new scenarios.  However, %additional observables may be required in order to determine the model parameters.

The remainder of this paper is organized as follows: In Section 2, we discuss the parameter space of this model and compare with the standard cosmology, followed by characterizing the SUSY signals at the LHC in Section 3.  In Section 4, we show the determination of model parameters and the prediction of relic density and neutralino-proton cross section. We conclude in Section 5, where some comments on the applicability of our main results to other string theory models are also discussed briefly.

%\newpage
%\section{Detection of  SSC at the LHC}
\section{Parameter Space}

\begin{figure}
\centering
\includegraphics[width=.70\textwidth]{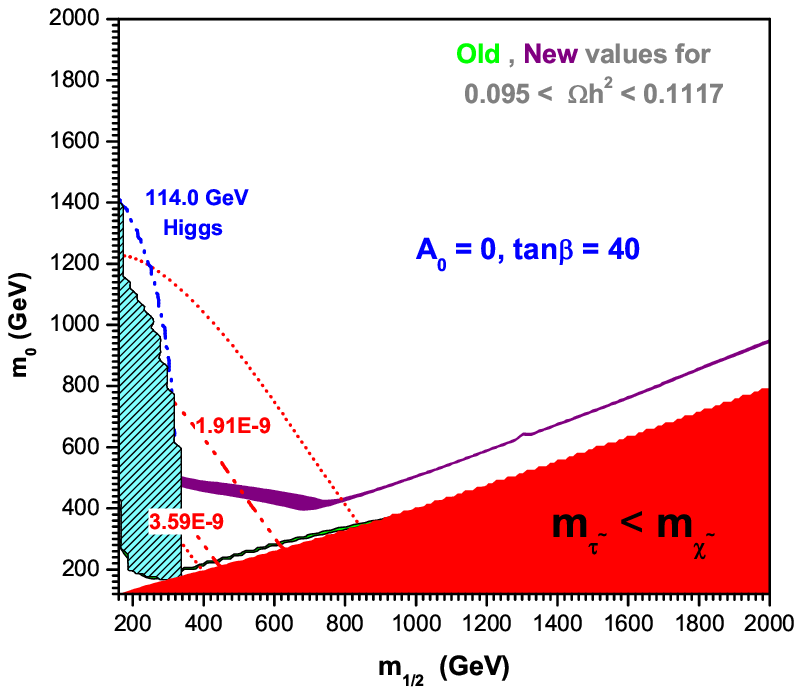}
\caption{WMAP allowed parameter space for the SSC and standard big bang cosmology shown for $\azero = 0$ $\tanb = 40$.  The very thin green (grey) band is where the neutralino relic density calculated by standard big bang cosmology agrees with the WMAP3 limits $0.0950 < \Omega  h^{2} < 0.1117$.  The thicker dark purple band shows the same agreement using the SSC calculation of the relic density.  Also shown are the Higgs mass boundary (dashed dotted blue line), muon $g_{\mu}-2$ boundaries (dashed and dotted red lines), a hatched cyan region which is excluded by $b \rightarrow s \gamma$ experimental bounds, and a lower solid red region where the neutralino is not the lightest supersymmetric particle.}
\label{figParameterSpace}
\end{figure}

The mSUGRA model parameters
are already significantly constrained by various experimental
results. Most important for limiting the parameter space are: (i)~the
light Higgs mass bound of $M_{h^0} > 114$~GeV from LEP~\cite{higgs1},
(ii)~the $b\rightarrow s \gamma$ branching ratio bound of
$1.8\times10^{-4} < {\cal B}(B \rightarrow X_s \gamma) <
4.5\times10^{-4}$ (we assume here a relatively broad range, since
there are theoretical uncertianties in extracting the branching ratio from
the data)~\cite{bsgamma}\footnote{The present experimental world average for $b\rightarrow s
\gamma$ is $(3.52 \pm 0.23\pm 9) \times 10{-4}$ \cite{bav} and the
SM contribution has been evaluated to be $(3.15\pm 0.23)\times
10{-4}$~\cite{misiak}. The $b\rightarrow s \gamma$ constraint does not have much of an impact in our analysis. The signals, Higgs+jets+ \met\ and 2
tau+jets+ \met\, in our study are available for a large region
of parameter space. The Z+jets+ \met\
final states however arise in the parameter space where
$b\rightarrow s \gamma$ is large which we mention later. We,
however, still discuss this final state since it is easy to
evade the $b\rightarrow s \gamma$ constraint without much change
in the final state.}, (iii)~the 2$\sigma$ bound on the dark matter
relic density: $0.095 < \Omega_{\rm CDM} h^2 <0.1117$~\cite{WMAP},
(iv)~the bound on the lightest chargino mass of $M_{\schionepm} >$
104~GeV from LEP~\cite{aleph} and (v) the muon magnetic moment anomaly
$a_\mu$, where the present deviation from the SM value
is $(29.5\pm 8.1) 10^{-10}$~\cite{raffael,BNL,dav,hag}. Assuming the future data
confirms the $a_{\mu}$ anomaly, the combined effects of $g_\mu -2$ and
$M_{\schionepm} >$ 104~GeV then only allows $\mu >0$.
Figure~\ref{figParameterSpace} shows g-2 curves for
$1.07\times 10^{-9}$, $1.91\times 10^{-9}$, $3.59\times10^{-9}$ and
$4.43\times10{-9}$ which  are within two sigma deviation.

Since the mSUGRA parameters determine the masses of our supersymmetric particles, they also determine the dark matter (or neutralino) relic density.  We can find which regions of the parameter space will agree with the dark matter relic density observed by WMAP.  The region allowed by WMAP for the standard big bang cosmology is vastly different from the region for the SSC model~\cite{ssc1}.  The comparison of these two regions is shown in Fig.~\ref{figParameterSpace} for the case of $\azero = 0$ and $\tanb = 40$.  We see a clear separation between the standard big bang cosmology region and the SSC region.  The standard big bang cosmology region is the very narrow region ranging from $350~\gev \lesssim \mhalf \lesssim 900~\gev$
($850~\gev \lesssim M_{\gluino} \lesssim 2000~\gev$), and $200~\gev \lesssim \mzero \lesssim 350~\gev$.  The SSC region is much broader for $\mhalf \lesssim 800~\gev$, and is higher in $\mzero$ which ranges from $400$ to $500~\gev$.
% Figure1

\section{Signals at the LHC}
\begin{figure}
\centering
\includegraphics[width=.70\textwidth]{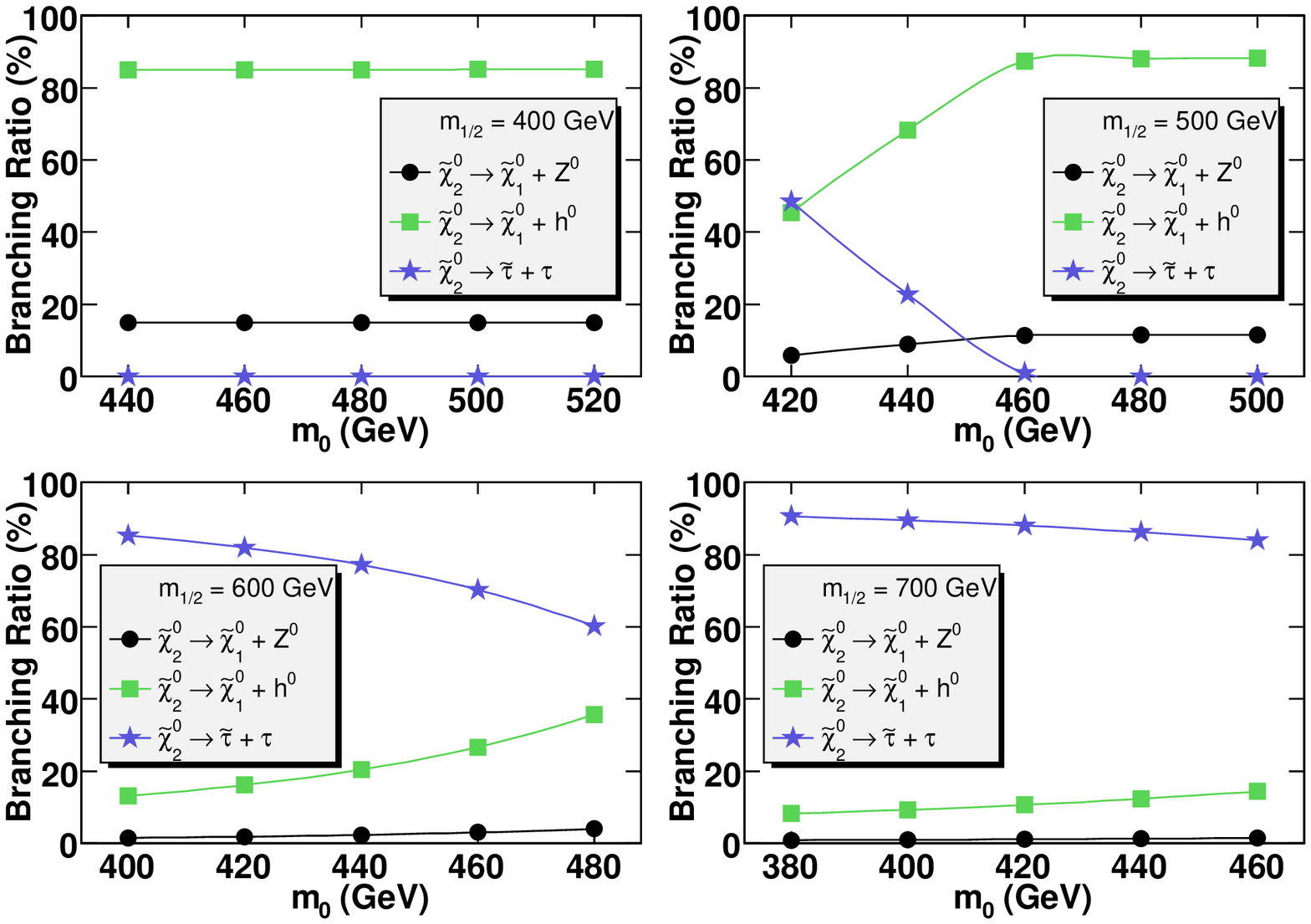}
\caption{The dominant decay branching ratios of decays from the $\schitwozero$ are shown here.  Each of the four plots shows how the branching ratios vary with $\mzero$ at constant $\mhalf$.  Together they survey the SSC band of parameter space in Fig.~\ref{figParameterSpace}.}
\label{figBranchingRatios}
\end{figure}

The SSC region of parameter space has some unique characteristics which are distinguishable from the $\stauone$-$\schionezero$ coannihilation region which appears for the lower values of $m_0$.
%In particular, for a large section of the SSC region, the particles we expect to see in abundance change.
For example, let us consider the decay chains of the dominant  SUSY production mechanism at the LHC, which will produce the squark and gluino, in pairs (e.g., $\squark \gluino$).  In the coannihilation region, the dominant decay chain for the squark ($\squarkL$) is $\squarkL \rightarrow q \schitwozero \rightarrow q \tau \stauone \rightarrow q \tau \tau \schionezero$.  Here $\schitwozero$ is the second lightest neutralino.  Thus, this region produces $\tau$'s, along with jets and missing transverse energy, $\met$.  However, the characteristic decay in the SSC region is $\schitwozero \rightarrow h^{0} \schionezero$.  In this case, we would expect $h^{0} \rightarrow b\bbar$, along with jets and $\met$.  Figure~\ref{figBranchingRatios} shows the branching ratios for the $\schitwozero$ decay.  As we increase $\mhalf$, the branching ratios shift from Higgs dominant decay chains to $\tau$ dominant decay chains.  However, we will easily distinguish this SSC $\tau$ dominated region fr
 om the coannihilation region by observing a large mass difference between the $\stauone$ and $\schionezero$.  For even lower $\mhalf$ values ($\mhalf \lesssim 350~\gev$), the $\schitwozero \rightarrow Z \schionezero$ decay becomes dominant.
%%%% Figure 2
Typical mass spectra are shown for points in the Higgs boson, $Z$ boson and two $\tau$ final state dominated regions in Tables~\ref{tabHiggsSignalSpectrum}, \ref{tabZSignalSpectrum}, and \ref{tabTauSignalSpectrum}.

\begin{table}
\caption{SUSY masses (in $\gev$) and dominant branching ratios for $\schitwozero$ for the point
$\mzero = 471~\gev$, $\mhalf = 440~\gev$, $\tanb = 40$, $\azero = 0$, and $\mu > 0$.  Notice that the $\schitwozero \rightarrow \stauone \tau$ is kinematically forbidden.  For this point, $\DMrelic = 0.089$ and $\pncross = 2.42\times10^{-9}\ \mathrm{pb}$.  The total production cross section for this point is $\sigma = 1.61~\mathrm{pb}$.}
 \label{tabHiggsSignalSpectrum}
\begin{center}
\begin{tabular}{c c c c c c c c}
\hline \hline
$\gluino$ &
$\begin{array}{c} \usquarkL \\ \usquarkR \end{array}$ &
$\begin{array}{c} \stoptwo \\ \stopone \end{array}$ &
$\begin{array}{c} \sbottomtwo \\ \sbottomone \end{array}$ &
$\begin{array}{c} \seleL \\ \seleR \end{array}$ &
$\begin{array}{c} \stautwo \\ \stauone \end{array}$ &
$\begin{array}{c} \schitwozero \\ \schionezero \end{array}$ &
$\begin{array}{c} \Br\left(\schitwozero\rightarrow h^{0} \schionezero\right)(\%) \\ \Br\left(\schitwozero\rightarrow Z \schionezero\right)(\%) \end{array}$
\\ \hline
1041 &
$\begin{array}{c} 1044 \\ 1017 \end{array}$ &
$\begin{array}{c} 954 \\ 768 \end{array}$ &
$\begin{array}{c} 958 \\ 899 \end{array}$ &
$\begin{array}{c} 557 \\ 500 \end{array}$ &
$\begin{array}{c} 532 \\ 393 \end{array}$ &
$\begin{array}{c} 341 \\ 181 \end{array}$ &
$\begin{array}{c} 86.8 \\ 13.0 \end{array}$
\\ \hline \hline
\end{tabular}
\end{center}
\end{table}

\begin{table}
\caption{SUSY masses (in $\gev$) and dominant branching ratios for $\schitwozero$ for the point
$\mzero = 471~\gev$, $\mhalf = 320~\gev$, $\tanb = 40$, $\azero = 0$, and $\mu > 0$.  We chose this point to examine despite the fact that it is within the region excluded by $b \rightarrow s \gamma$.  We did this to examine the behavior of $\schitwozero\rightarrow Z \schionezero$ at its maximal branching ratio.  The total production cross section for this point is $\sigma = 7.10~\mathrm{pb}$.}
 \label{tabZSignalSpectrum}
\begin{center}
\begin{tabular}{c c c c c c c c}
\hline \hline
$\gluino$ &
$\begin{array}{c} \usquarkL \\ \usquarkR \end{array}$ &
$\begin{array}{c} \stoptwo \\ \stopone \end{array}$ &
$\begin{array}{c} \sbottomtwo \\ \sbottomone \end{array}$ &
$\begin{array}{c} \seleL \\ \seleR \end{array}$ &
$\begin{array}{c} \stautwo \\ \stauone \end{array}$ &
$\begin{array}{c} \schitwozero \\ \schionezero \end{array}$ &
$\begin{array}{c} \Br\left(\schitwozero\rightarrow h^{0} \schionezero\right)(\%) \\ \Br\left(\schitwozero\rightarrow Z \schionezero\right)(\%) \end{array}$
\\ \hline
785 &
$\begin{array}{c} 838 \\ 821 \end{array}$ &
$\begin{array}{c} 763 \\ 598 \end{array}$ &
$\begin{array}{c} 768 \\ 708 \end{array}$ &
$\begin{array}{c} 519 \\ 487 \end{array}$ &
$\begin{array}{c} 493 \\ 389 \end{array}$ &
$\begin{array}{c} 241 \\ 129 \end{array}$ &
$\begin{array}{c} 0. \\ 99.6 \end{array}$
\\ \hline \hline
\end{tabular}
\end{center}
\end{table}

\begin{table}
\caption{SUSY masses (in $\gev$) and dominant branching ratios for $\schitwozero$ for the point
$\mzero = 440~\gev$, $\mhalf = 600~\gev$, $\tanb = 40$,
$\azero = 0$, and $\mu > 0$.  For this point, $\DMrelic = 0.106$ and $\pncross = 7.19\times10^{-10}\ \mathrm{pb}$.  The total production cross section for this point is $\sigma = 0.446~\mathrm{pb}$.}
 \label{tabTauSignalSpectrum}
\begin{center}
\begin{tabular}{c c c c c c c c}
\hline \hline
$\gluino$ &
$\begin{array}{c} \usquarkL \\ \usquarkR \end{array}$ &
$\begin{array}{c} \stoptwo \\ \stopone \end{array}$ &
$\begin{array}{c} \sbottomtwo \\ \sbottomone \end{array}$ &
$\begin{array}{c} \seleL \\ \seleR \end{array}$ &
$\begin{array}{c} \stautwo \\ \stauone \end{array}$ &
$\begin{array}{c} \schitwozero \\ \schionezero \end{array}$ &
$\begin{array}{c} \Br\left(\schitwozero\rightarrow h^{0} \schionezero\right)(\%) \\ \Br\left(\schitwozero\rightarrow \tau \stauone \right)(\%) \end{array}$
\\ \hline
1366 &
$\begin{array}{c} 1252 \\ 1211 \end{array}$ &
$\begin{array}{c} 1153 \\ 957 \end{array}$ &
$\begin{array}{c} 1153 \\ 1094 \end{array}$ &
$\begin{array}{c} 594 \\ 494 \end{array}$ &
$\begin{array}{c} 574 \\ 376 \end{array}$ &
$\begin{array}{c} 462 \\ 249 \end{array}$ &
$\begin{array}{c} 20.5 \\ 77.0 \end{array}$
\\ \hline \hline
\end{tabular}
\end{center}
\end{table}

\subsection{Signals}

We have three possible signals in this model. These signals are the following: \begin{itemize}
\item Higgs + jets + $\met$
\item $Z$ + jets + $\met$
\item 2 $\tau$ + jets + $\met$
\end{itemize}

\subsubsection{Higgs + Jets + $\met$}
The Higgs + jets + $\met$ signal appears in the lower $\mhalf$ region ($400 \lesssim \mhalf \lesssim500~\gev$) within the SSC band of parameter space in Fig.~\ref{figParameterSpace}.   The Higgs + jets + $\met$ signal is characterized by the decay chain, $\squarkL \rightarrow q \schitwozero \rightarrow q h^{0} \schionezero$.  The $\schionezero$ does not interact in the detector, and thus leaves a large $\met$ signal.  The $h^{0}$ and jet carry information about the SUSY particles in this chain.  In particular, the $h^{0}$ + jet invariant mass distribution has an endpoint which depends upon their masses:
\begin{equation}
M_{h^{0} q}^{\mathrm{end}} = \sqrt{M_{h^{0}}^{2} + \frac{\left( M_{\squarkL}^{2} - M_{\schitwozero}^{2}  \right) \left(M_{\schitwozero}^{2} + M_{h^{0}}^{2} - M_{\schionezero}^{2} + \sqrt{(M_{\schitwozero}^{2} - M_{h^{0}}^{2} - M_{\schionezero}^{2})^{2} - 4M_{h^{0}}^{2} M_{\schionezero}^{2}}\right)} { 2M_{\schitwozero}^{2}}}
\label{eqEndpointHiggsPlusJet}
\end{equation}
For our analysis, we generate events using \pythia~\cite{pythia}, which is linked with \isasugra~\cite{isajet} to generate the mSUGRA mass spectrum.  We pass these events to a detector simulator called \pgs~\cite{pgs}.

To measure the endpoint, we begin by selecting our events with the following cuts~\cite{LHCtwotau}:\begin{itemize}
\item At least two jets with $p_{T} \ge 200~\gev$ as well as $|\eta| \le 2.5$,
\item $\met \ge 180~\gev$,
\item $p_{T}^{\mathrm{jet1}} + p_{T}^{\mathrm{jet2}} + \met \ge 600~\gev$, and
\item at least two $b$-tagged jets~\cite{pgs} with $p_{T} \ge 100~\gev$ and $|\eta| \le 1.5$.
\end{itemize}  These cuts remove the majority of the SM background, such as $t\tbar$, $W$+jets, and $Z$+jets~\cite{hinch1}, as well as some background from SUSY events which do not contain the decay chain $\squarkL \rightarrow q \schitwozero \rightarrow q h^{0} \schionezero$.  Unforeseen SM backgrounds at the LHC can be shape analyzed from the data and then subtracted from our desired signal.

\begin{figure}[t]
\centering
\includegraphics[width=.70\textwidth]{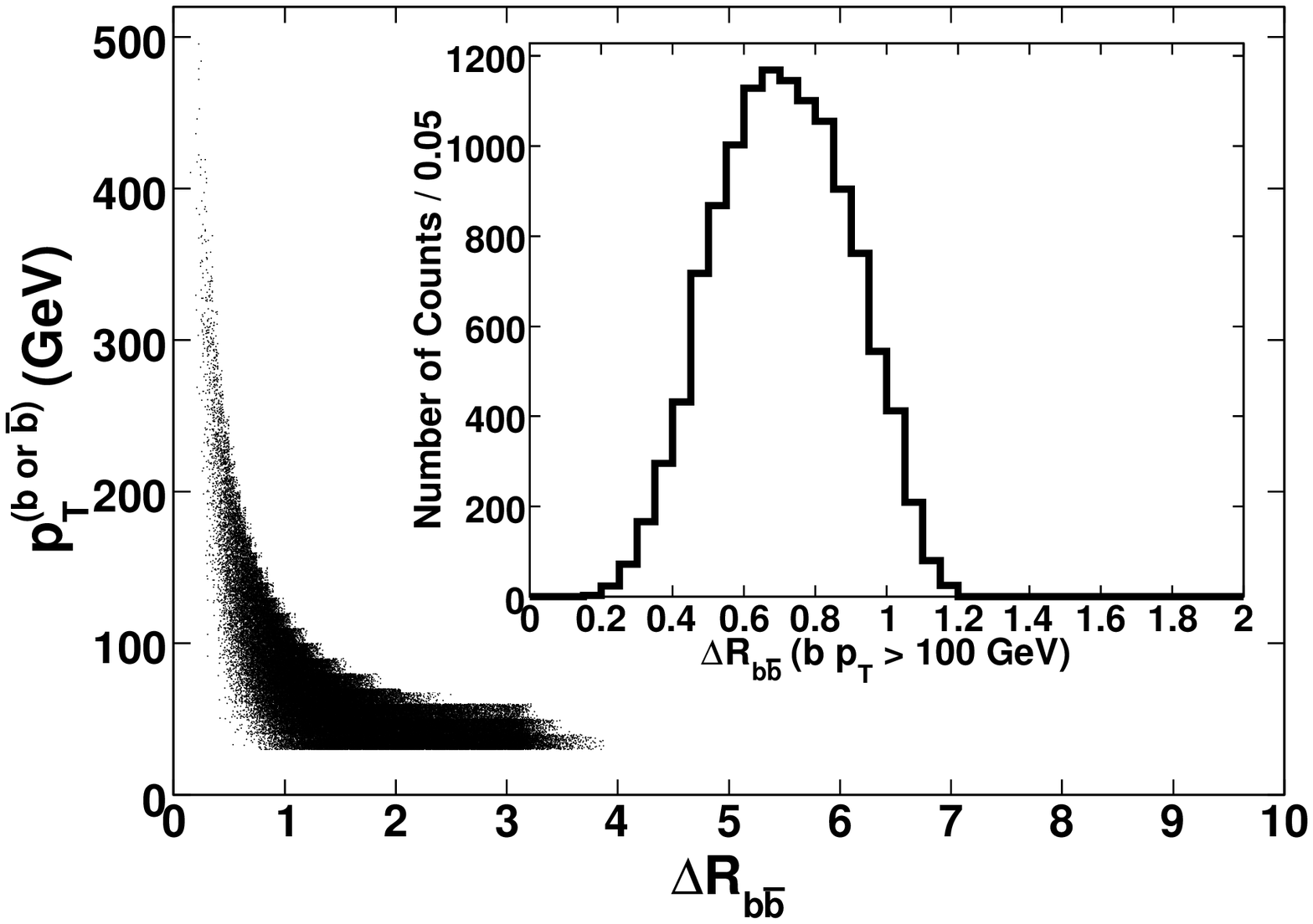}
\caption{Correlation between $p_{T}^{\rm min} \equiv \min(p_{T}^{b}, p_{T}^{\bar b})$ and $\Delta R_{\bbbar}$ from $h^{0}\rightarrow \bbbar$ at the generator level for mSUGRA events at $\mzero = 471~\gev$, $\mhalf = 400~\gev$.  Here $\Delta R_{\bbbar}$ is a separation between $b$ and $\bbar$ in $\eta$-$\phi$ space.  The inset histogram is the $\Delta R_{\bbbar}$ distribution for $p_{T}^{\rm min} > 100~\gev$.  This shows that $\bbbar$ pairs from a single Higgs decay will most often have a separation of $\Delta R < 1$ for $b$-jets with transverse momentum greater than $100~\gev$.  Any $b$ pairs not from a single Higgs decay will instead have no particular separation.}
\label{figHiggsToBBTruth}
\end{figure}

\begin{figure}[t]
\centering
\includegraphics[width=.70\textwidth]{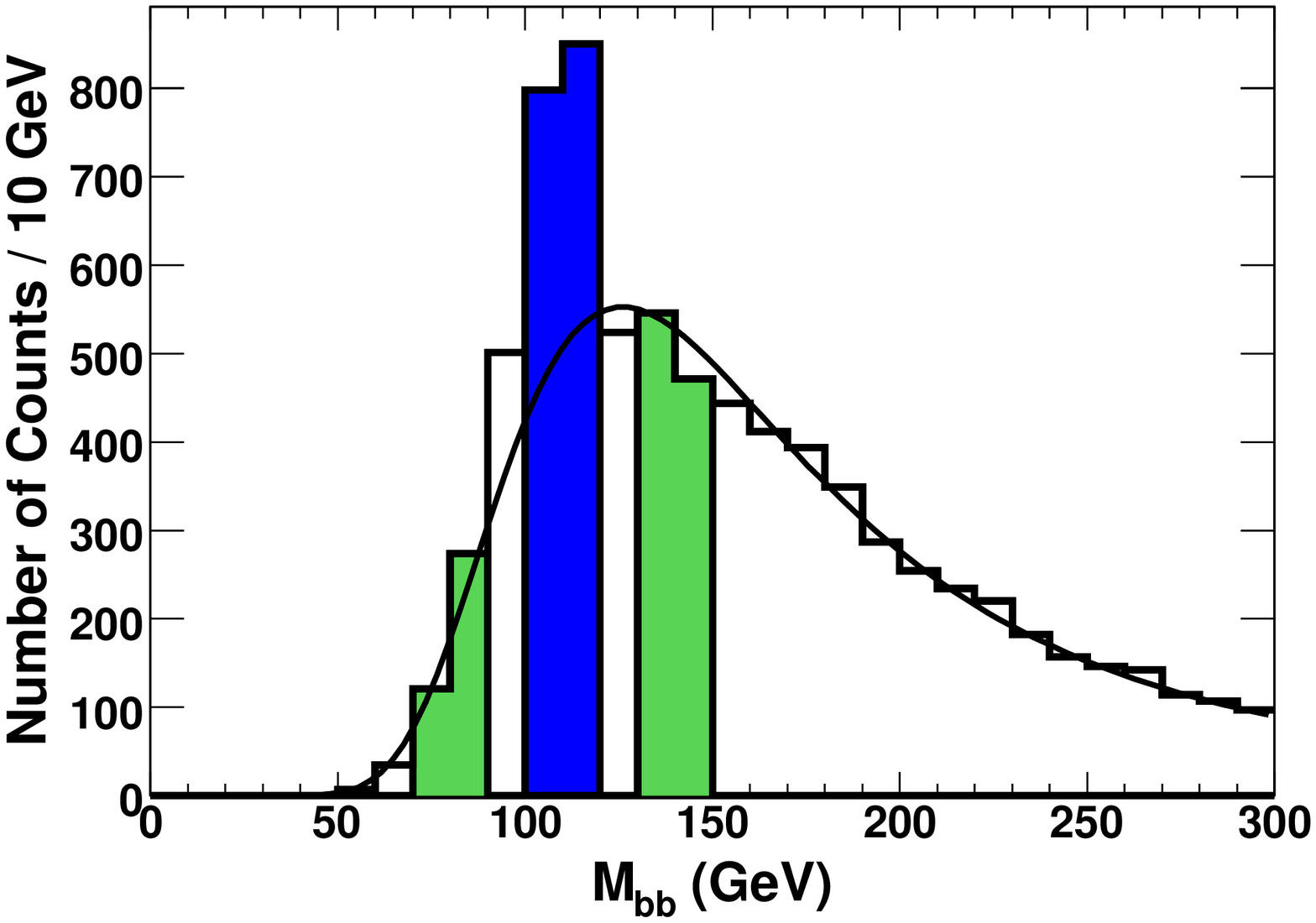}
\caption{The invariant mass distribution of PGS $b$ jet pairs.  The central blue (dark grey) bins are the Higgs peak window.  We perform a background subtraction by selecting the green (light grey) sideband windows.  The background fit is the black curve.}
\label{figDibMass}
\end{figure}

Next we identify Higgs bosons in the event. We select all pairs of $b$-tagged jets with $p_{T}^{b}\ge100~\gev$ and $0.4 < \Delta R_{bb} < 1$.  The lower $\Delta R$ limit is due to a jet clustering cone size in \pgs, while the upper limit is motivated by a study of Higgs decays in mSUGRA events ($\mzero = 471~\gev$, $\mhalf = 400~\gev$) at the generator level. (See Fig.~\ref{figHiggsToBBTruth}.)  We then form the $b$ pair invariant mass.  Figure~\ref{figDibMass} shows a peak between $100~\gev$ and $120~\gev$, consistent with the Higgs mass, along with a continuum background.
% Figure 3
% Figure 4
For each $b$ pair, we form $2b+\mathrm{jet}$ systems, using the two jets with the greatest transverse momenta of the event.   These two leading jets will primarily come from three different decay chains:\begin{itemize}
\item $\squark \rightarrow q \gluino$
\item $\squarkR \rightarrow q \schionezero$
\item $\squarkL \rightarrow q \schitwozero \rightarrow q h^{0} \schionezero$
\end{itemize}
Each $b$ pair combined with these two leading jets will form two effective masses, $M_{bbj_{1}}$, and $M_{bbj_{2}}$.  If we combine the jet from $\squarkR \rightarrow q \schionezero$ with our Higgs, it can have a larger $M_{bbj}$ than the endpoint expected from the $\squarkL \rightarrow q \schitwozero \rightarrow q h^{0} \schionezero$ decay chain.  Thus, we simply choose the lesser of $M_{bbj_{1}}$ and $M_{bbj_{2}}$, denoting it as $M_{bbj}^{\mathrm{2nd}}$.  This selection is a similar to that shown in Ref.~\cite{hinch1}.

At this stage, we still suffer from $bb$ combinatoric background as seen in Fig.~\ref{figDibMass}. To estimate the combinatoric background in the Higgs mass window, we perform a sideband subtraction method.  We form the $M_{bbj}^{\mathrm{2nd}}$ distribution using $b$ pairs in the Higgs peak window  and using $b$ pairs from two sideband windows ($70$-$90~\gev$ and $130$-$150~\gev$) in Fig.~\ref{figDibMass}.  This second distribution is scaled by the ratio of the background shape evaluated in the Higgs window to the sideband windows.  (See Fig.~\ref{figDibMass}).  Then we subtract this scaled sideband $M_{bbj}^{\mathrm{2nd}}$ distribution from the Higgs $M_{bbj}^{\mathrm{2nd}}$ distribution.  Since the kinematical endpoint occurs when the Higgs is back to back with the jet, we select events with $\Delta R_{h^{0}j} > 1.2$.

\begin{figure}
\centering
\includegraphics[width=.70\textwidth]{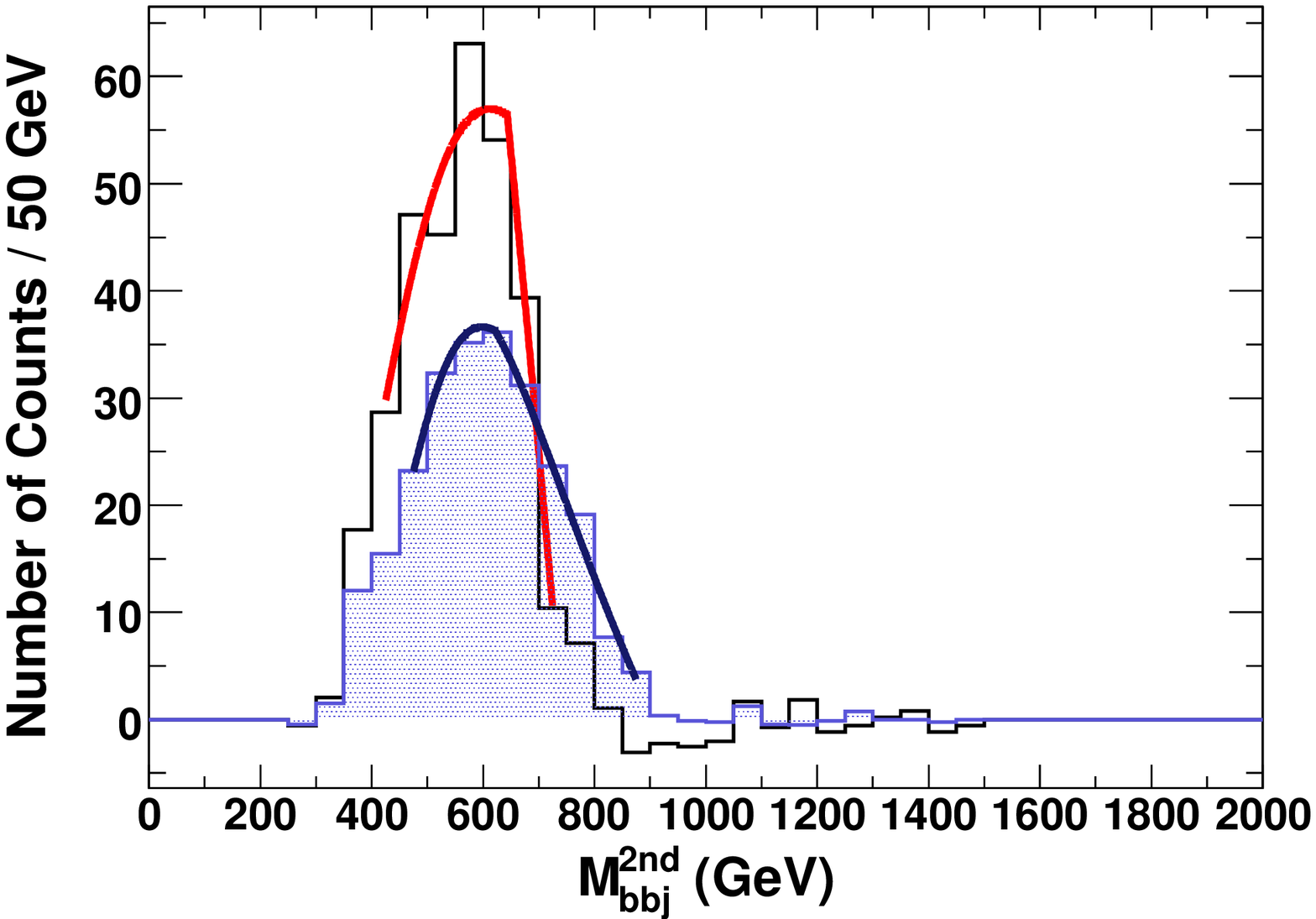}
\caption{The Higgs (tagged $b$ jet pair) plus jet invariant mass distribution reconstructed through PGS in two $500~\invfb$ mSUGRA samples at $(\mzero,~\mhalf) = (471~\gev,~400~\gev)$ and $(471~\gev,~480~\gev)$ for the black histogram with red (gray) fit and blue (gray, filled) histogram with dark blue(dark gray) fit, respectively.  We fix $\tanb = 40$, $\azero = 0$, and $\mu > 0$.}
\label{figHiggsPlusJetLandauCompare}
\end{figure}

In order to determine the endpoint, we fit the mass distribution to a combination of a Landau probability distribution function, $P_{L}$, and a straight line:
\begin{equation}
f(x)=\left\{\begin{array}{c c} kP_{L}(x, x_{peak}, \sigma) & \mathrm{if}\  x < x_{peak} \\ kP_{L}(x, x_{peak}, \sigma) + \alpha(x-x_{peak}) & \mathrm{if}\  x > x_{peak} \end{array}\right.,
\label{eqLandauPlusLine}
\end{equation}
where $x$ corresponds to the Higgs plus jet invariant mass, $x_{peak}$ is the most probable value of the Landau distribution, $k$ scales the height of the function, and $\alpha$ is the slope of the linear portion.  Figure~\ref{figHiggsPlusJetLandauCompare} shows the fittings at two mSUGRA points around our reference point described in Table~\ref{tabHiggsSignalSpectrum}. One can see the change in the shape and the end point as $\mhalf$ increases.  The slight change in shape between the two histograms in Fig.~\ref{figHiggsPlusJetLandauCompare} is due to the fact that the SUSY background for this signal does not shift as we vary $\mhalf$, and that it dies off around $800~\gev$.  Thus, the $\mhalf = 400~\gev$ histogram which has an endpoint around $750~\gev$ has a slight shoulder after the endpoint, whereas the $\mhalf = 480~\gev$ histogram has an endpoint around $900~\gev$ with no shoulder.

\begin{figure}
\centering
\includegraphics[width=.70\textwidth]{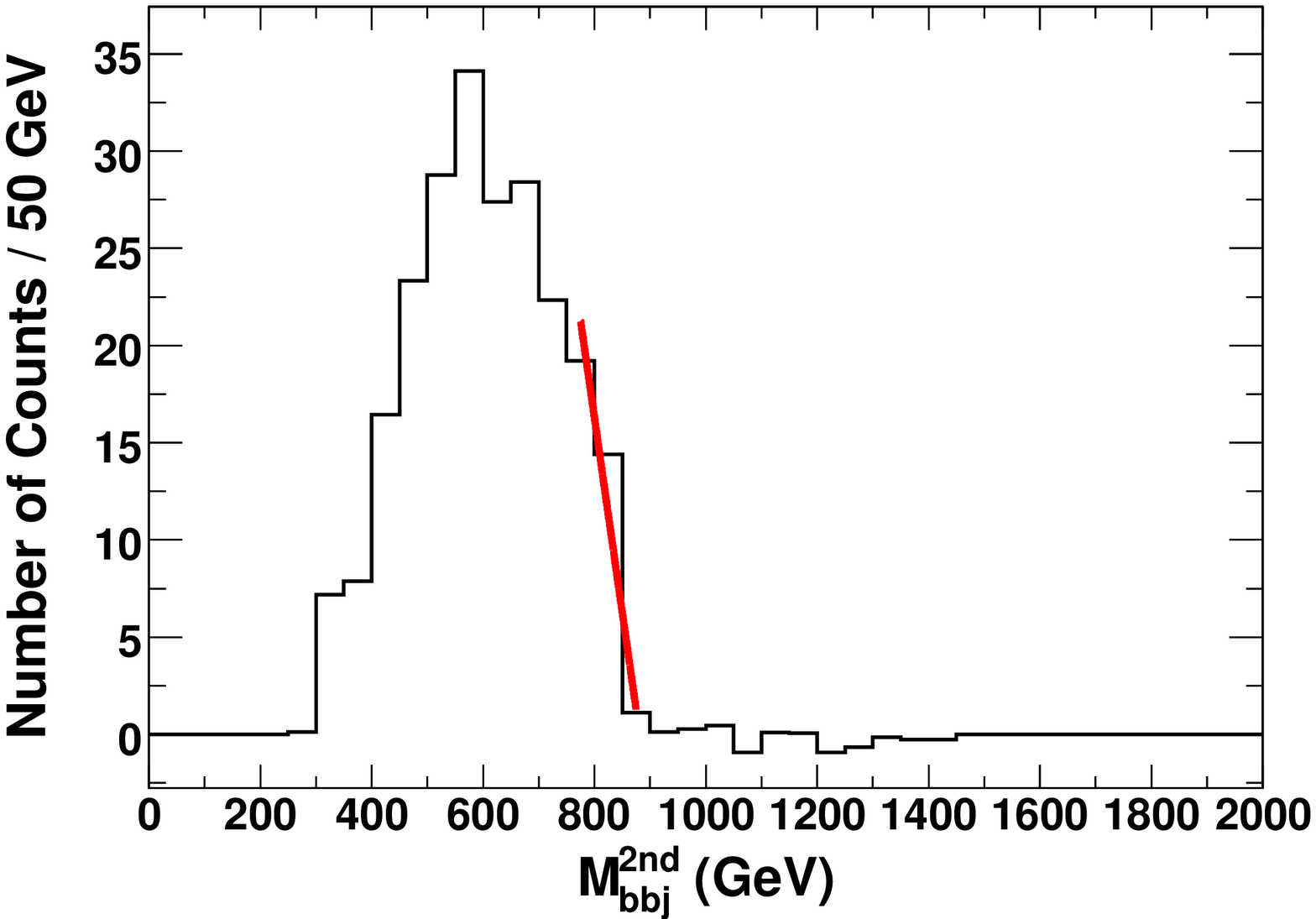}
\caption{The Higgs (tagged $b$ jet pair) plus jet invariant mass distribution reconstructed through PGS in a $500~\invfb$ mSUGRA sample at $\mzero = 651~\gev$, $\mhalf = 440~\gev$, $\tanb = 40$, $\azero = 0$, and $\mu > 0$. }
\label{figHiggsPlusJet651440}
\end{figure}

Despite such shoulders, the shape of the distribution stays similar as we vary the model parameter $\mhalf$.  Thus we can use the same fitting function for such points.  Also, since the endpoint, $M_{h^{0} q}^{\mathrm{end}}$ does not depend on any third generation sparticles (see Eq.~\ref{eqEndpointHiggsPlusJet}), it is independent of the parameters $\azero$ and $\tanb$.  However, if we increase $\mzero$ the situation changes.  For higher $\mzero$ values the $\squark$ becomes significantly heavier than the $\gluino$.  The result of this, for instance in the case of $\mzero = 651~\gev$ and $\mhalf = 440~\gev$, is that $\Br(\squarkL \rightarrow q\gluino) = 10\%$ and $\Br(\squarkR \rightarrow q\gluino) = 22\%$.  The quark jets from such decay chains are much softer than background jets from lower $\mzero$ points.  Thus, the nature of the background changes, which changes the shape of the Higgs plus jet invariant mass distribution:  The endpoint becomes very sharp.  Thus a simple
  linear fit is sufficient to find the endpoint.  A sample fit of this higher $\mzero$ region is shown in Fig.~\ref{figHiggsPlusJet651440}.
% Figure 5
% Figure 6

\subsubsection{$Z$ + Jets + $\met$}
The final state of $Z$ + jets + $\met$ events becomes a key signal in a lower $\mhalf$ region  ($\mhalf\simeq300~\gev$) where the $\schitwozero \rightarrow q h^{0} \schionezero$ decays are kinematically suppressed.  The decay chain and endpoint equation (Eq.~\ref{eqEndpointHiggsPlusJet}) are exactly the same under the replacement of the Higgs boson mass with the $Z$ boson mass.

\begin{figure}
\centering
\includegraphics[width=.70\textwidth]{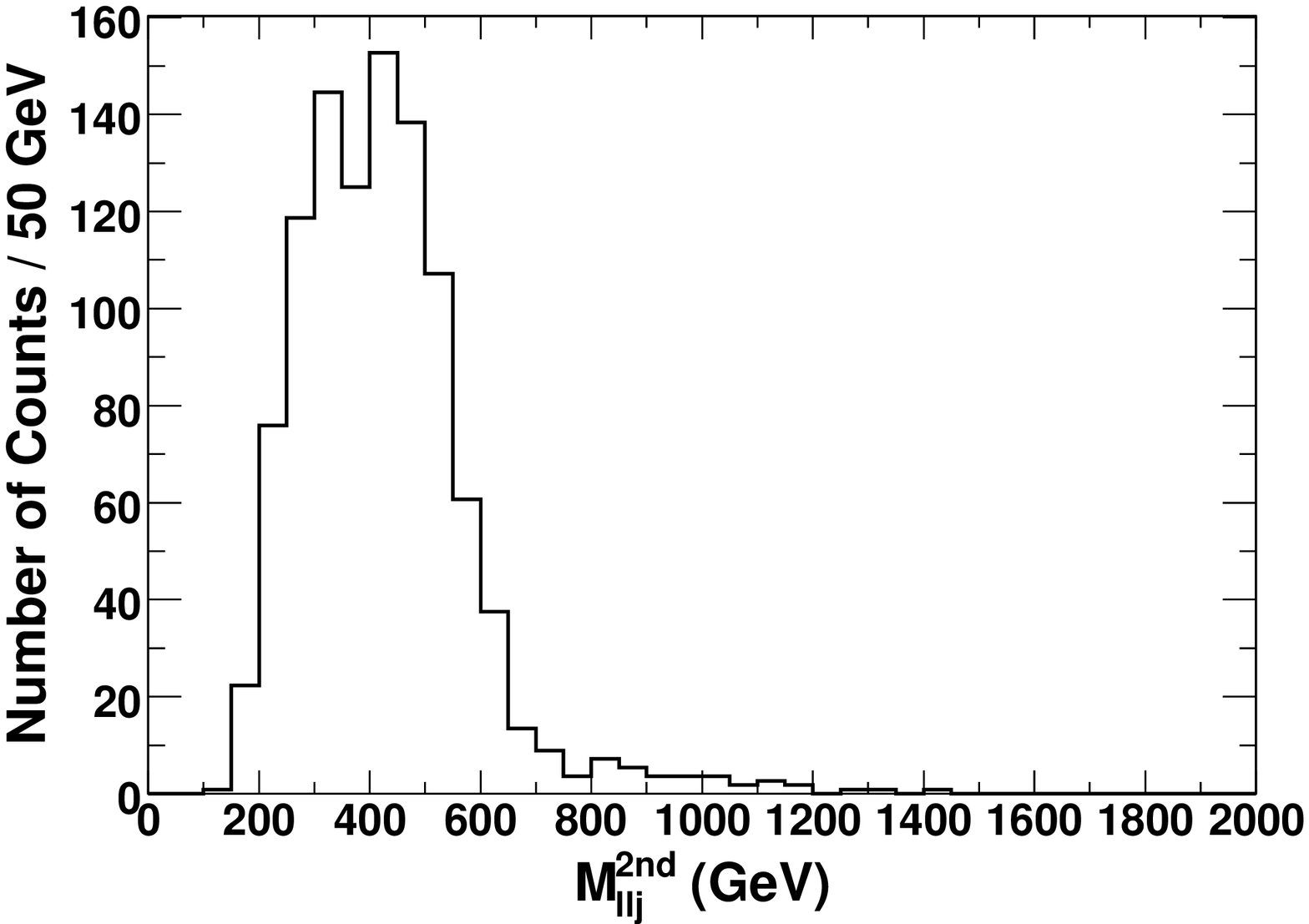}
\caption{The $Z$ plus jet invariant mass distribution reconstructed through PGS in one $50~\invfb$ mSUGRA sample at $\mzero = 471~\gev$, $\mhalf = 320~\gev$, $\tanb = 40$, $\azero = 0$, and $\mu > 0$.}
\label{figZPlusJet471320}
\end{figure}

To construct the $Z$ plus jet invariant mass and measure the endpoint, we follow the very same procedure as the Higgs plus jet analysis, but with $Z \to ll$ decays.  We select events with the same initial cuts as in the Higgs plus jet analysis and reconstruct the $Z \to ll$ decays instead of the Higgs decays.  To select our $Z$ bosons we find pairs of isolated leptons with $p_{T} > 20~\gev$ in our event.  We keep pairs of leptons with invariant mass within the $Z$ mass window, where $85~\gev < M_{ll} < 97~\gev$.  Then we form $2l+\mathrm{jet}$ systems using the two jets with the greatest transverse momenta.  We again keep only the lesser of the two $2l+\mathrm{jet}$ invariant masses, $M_{llj}^{\mathrm{2nd}}$.  To ensure we select mostly $Z$ bosons within this signal we use an opposite-sign-same-flavor minus opposite-sign-opposite-flavor subtraction.  A sample distribution of the result is shown in Fig.~\ref{figZPlusJet471320}.  This figure shows us a measurable endpoint very
 similar to that of the Higgs plus jet invariant mass technique.

%%%% Figure 7

\subsubsection{2$\tau$ + Jets + $\met$}

%%%% Table 3

The 2$\tau$ + jets + $\met$ signal appears in the higher $\mhalf$ region ($\mhalf \gtrsim 500~\gev$) within the SSC band of parameter space in Fig.~\ref{figParameterSpace}.  The full decay chain $\squarkL \rightarrow q \schitwozero \rightarrow \stauone \tau \rightarrow \tau \tau \schionezero$ produces a characteristic final state consisting of $\tau$'s, high $p_{T}$ jets from the $\squarkL$ decay, and $\met$ from the $\schionezero$.  Again, the two $\tau$ particles and the jet carry information about the supersymmetric particles in the decay chain.  The visible ditau invariant mass distribution and the $2\tau$ plus jet invariant mass distribution both have endpoints depending on the supersymmetric particle masses:
\begin{equation}
M_{\tau\tau}^{\mathrm{end}}=M_{\schitwozero}\sqrt{\left(1-\frac{M_{\stauone}^2}{M_{\schitwozero}^2}\right)\left(1-\frac{M_{\schionezero}^2}{M_{\stauone}^2}\right)}
\label{eqEndpointTwoTau}
\end{equation}
\begin{equation}
M_{j\tau\tau}^{\mathrm{end}}=M_{\squarkL}\sqrt{\left(1-\frac{M_{\schitwozero}^2}{M_{\squarkL}^2}\right)\left(1-\frac{M_{\schionezero}^2}{M_{\schitwozero}^2}\right)}
\label{eqEndpointJetTauTau}
\end{equation}
We again make use of such kinematical observables by measuring these endpoints.  However, in this case, we are restricted by the background for $M_{\tau\tau}^{\mathrm{end}}$.  To avoid this background, we measure the peak instead, since the peak is proportional to the endpoint.  Events were generated using \isajet~\cite{isajet} and the detector effects were simulated using \pgs~\cite{pgs}.

To measure these observables, we select our events with the following cuts~\cite{LHCrelicdensity}:\begin{itemize}
\item At least two jets with $p_{T} \ge 200~\gev$ as well as $|\eta| \le 2.5$,
\item $\met \ge 180~\gev$, and
\item $p_{T}^{\mathrm{jet1}} + p_{T}^{\mathrm{jet2}} + \met \ge 600~\gev$.
\item We reject events where either one of the two leading jets is tagged as a $b$ jet~\cite{pgs}.
\end{itemize}  These cuts remove the majority of SM backgrounds ($t\tbar$, $W$+jets, and $Z$+jets), as well as background from SUSY events containing stops or sbottoms.  Here again, unforeseen SM backgrounds at the LHC can be shape analyzed from the data and then subtracted from our desired signal.  We do not discuss the details of the event selections, but instead refer the reader to our previous and ongoing studies~\cite{LHCtwotau, LHCthreetau, LHCrelicdensity}.

\begin{figure}
\centering
\includegraphics[width=.70\textwidth]{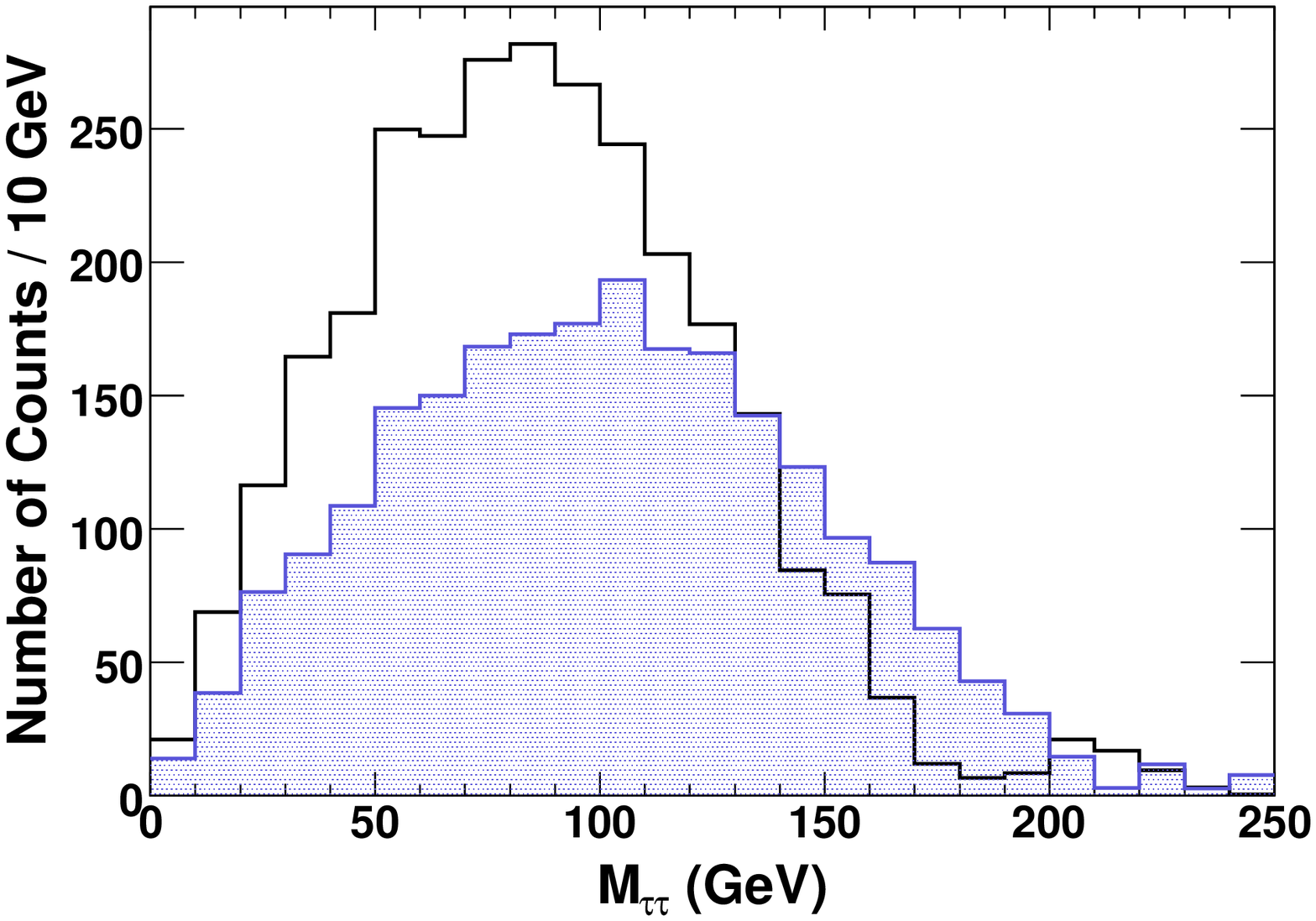}
\caption{The ditau invariant mass distribution reconstructed through PGS in two $500~\invfb$ mSUGRA samples at $(\mzero,~\mhalf) = (440~\gev,~625~\gev)$ and $(440~\gev,~575~\gev)$ for the black and blue(gray, filled) histograms, respectively.  We fix $\tanb = 40$, $\azero = 0$, and $\mu > 0$.}
\label{figMtautauCompare}
\end{figure}

\begin{figure}[t]
\centering
\includegraphics[width=.70\textwidth]{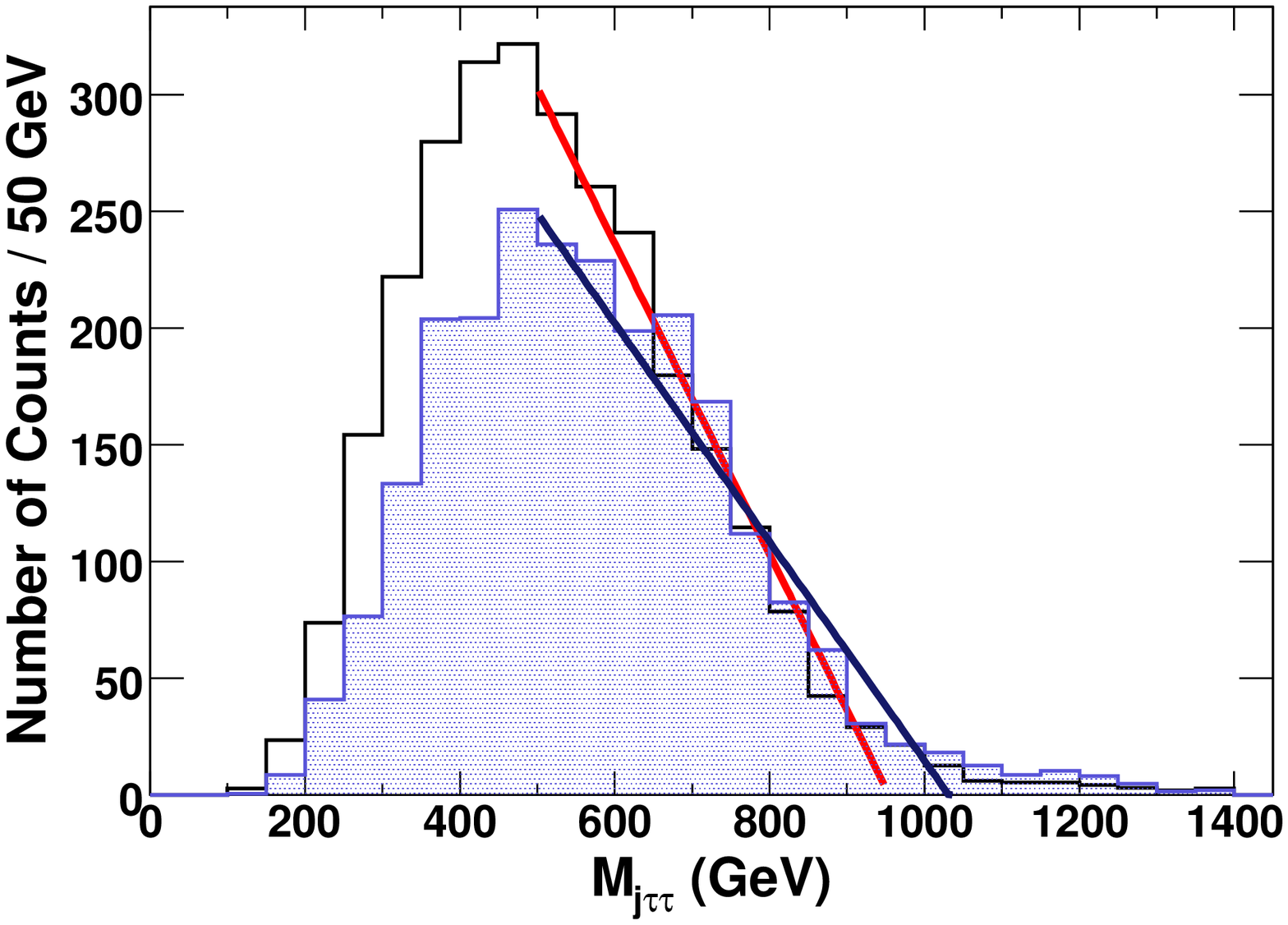}
\caption{The $2\tau$ plus jet invariant mass distribution reconstructed through PGS in two $500~\invfb$ mSUGRA samples at $(\mzero,~\mhalf) = (440~\gev,~625~\gev)$ and $(440~\gev,~575~\gev)$ for the black histogram with red(gray) fit and blue(gray, filled) histogram with dark blue(dark gray) fit, respectively.  We fix $\tanb = 40$, $\azero = 0$, and $\mu > 0$.}
\label{figMjettautauCompare}
\end{figure}

Sample $M_{\tau\tau}$ and $M_{j\tau\tau}$ distributions for points similar to that shown in Table~\ref{tabTauSignalSpectrum} are displayed in Figs.~\ref{figMtautauCompare} and \ref{figMjettautauCompare}.  These figures also show how the peak and endpoint shift under changes of $\mhalf$.  Since the endpoint of the $2\tau$ + jet invariant mass distribution, $M_{j\tau\tau}^{\mathrm{end}}$, does not depend on any third generation superparticles (see Eq.~\ref{eqEndpointJetTauTau}), it will only shift for variations of the mSUGRA parameters $\mzero$ and $\mhalf$.  However, the peak of the $2\tau$ invariant mass distribution, $M_{\tau\tau}^{\mathrm{peak}}$, depends on the stau mass (see Eq.~\ref{eqEndpointTwoTau}), and will thus depend on all four mSUGRA parameters.

%%%% Figure 8

%%%% Figure 9

%%%%

\section{Determining Model Parameters}

We have shown in previous works~\cite{LHCtwotau, LHCthreetau} that we can obtain mass measurements of the supersymmetric particles in the neutralino-stau coannihilation region  by utilizing each final state and parameterizing kinematical observables, such as those described in the previous section, in terms of the SUSY masses.  Our goal is to determine the mSUGRA model parameters $\mzero$, $\mhalf$, $\azero$, and $\tanb$ since we want to determine the dark matter content and the neutralino-proton cross section.  The fifth mSUGRA model parameter, $\mathrm{sign}(\mu)$, is assumed to be positive, since this is preferred by measurements of the $b\rightarrow s\gamma$ branching ratio and the muon $g_{\mu}-2$~\cite{BNL}.  To determine the mSUGRA parameters, we will thus need four kinematical observables which are linearly independent functions of those parameters.  The determination of the parameters is then accomplished by inverting four such functional relationships.

As discussed above, certain regions of the mSUGRA parameter space might give rise to very different signals. For each region we can combine different observables to determine the four mSUGRA parameters.  However, so far, there are not four observables which can be made for each signal described above.  Thus we introduce the following additional kinematical observables which are valid in any region: $\meff, \meffb,$ and $\mefftwob$.

The effective mass, $\meff$, is defined by
\begin{equation}
\meff = p_{T}^{\mathrm{jet1}} + p_{T}^{\mathrm{jet2}} + p_{T}^{\mathrm{jet3}} + p_{T}^{\mathrm{jet4}} + \met,
\label{eqMeffective}
\end{equation}
where all four leading jets are not $b$-tagged jets.
This combination carries the information of the characteristic SUSY scale.  The majority of the $\pt$ of the jets is characteristic of the gluino and first two generation squark decays, and the majority of the $\met$ is due to the lightest neutralino escaping the detector.  As such, the observable $\meff$ depends only on the mSUGRA parameters $\mzero$ and $\mhalf$.  This is because the parameters $\azero$ and $\tanb$ only affect the third generation superparticles.

\begin{figure}[t]
\centering
\includegraphics[width=.70\textwidth]{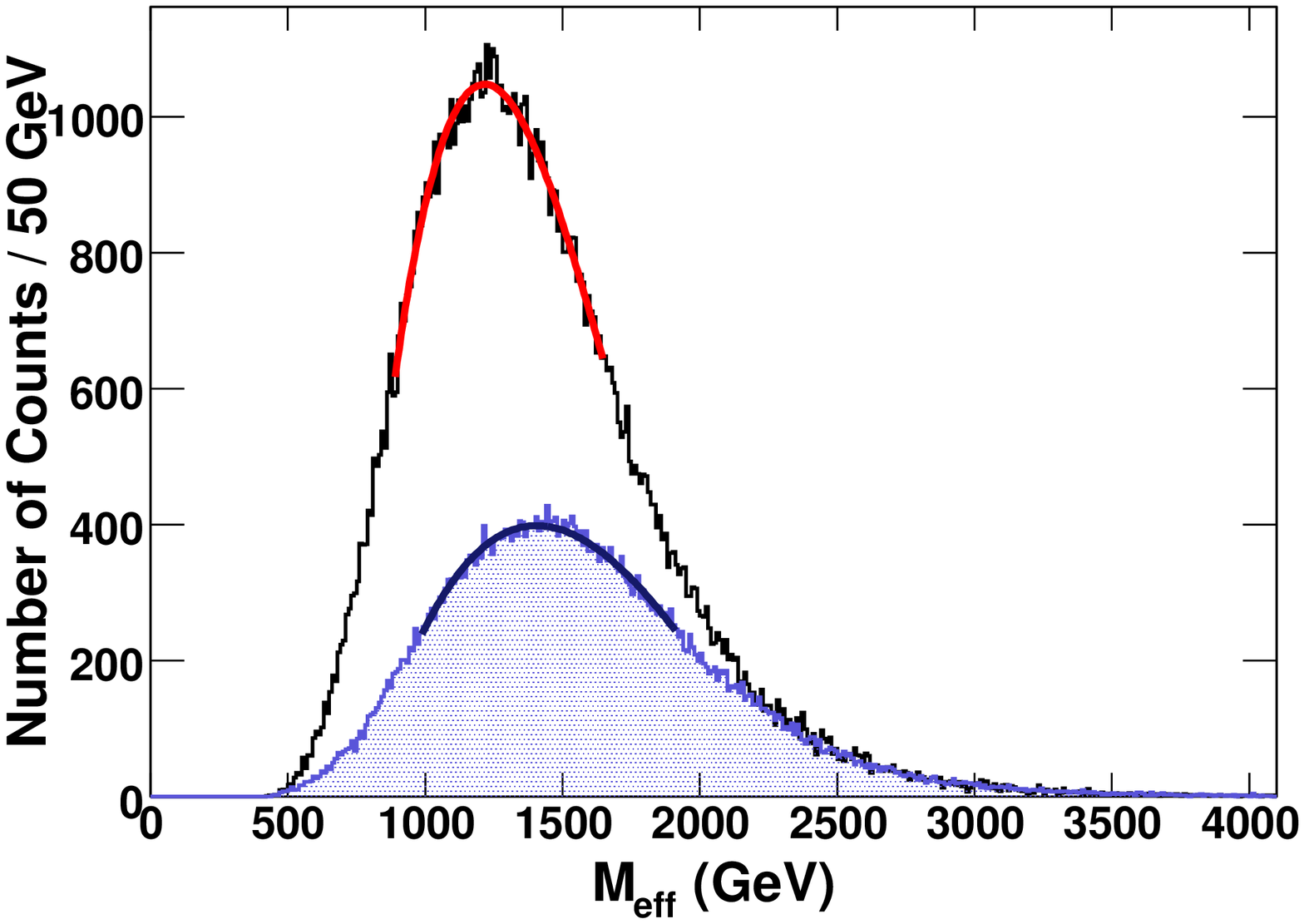}
\caption{The effective mass distribution reconstructed through PGS in two $500~\invfb$ mSUGRA samples at $(\mzero,~\mhalf) = (471~\gev,~400~\gev)$ and $(471~\gev,~480~\gev)$ for the black histogram with red(gray) fit and blue(gray, filled) histogram with dark blue(dark gray) fit, respectively.  We fix $\tanb = 40$, $\azero = 0$, and $\mu > 0$.}
\label{figMeffectiveCompare}
\end{figure}

When we construct the effective mass distribution, we use the following cuts\cite{hinch1}:\begin{itemize}
\item At least one jet with $p_{T} \ge 100~\gev$ and an additional three jets $p_{T} \ge 50~\gev$, where all such jets have $|\eta| \le 2.5$,
\item No isolated leptons with $|\eta| \le 2.5$,
\item $\met \ge 200~\gev$ and $\met \ge 0.2 \times \meff$,
\item Transverse Sphericity, $S_{T} > 0.2$,
\item We reject events where any of the four leading jets is tagged as a $b$ jet\cite{pgs}.
\end{itemize}  We find the peak value with an iterative fitting technique.  First, we fit the distribution iteratively with an asymmetric gaussian function.  The purpose of the iterative fit is simply to find the ideal fitting range.  Once that is found, we fit with a cubic polynomial to find the peak position.  A sample effective mass distribution showing the result of this procedure is shown in Fig.~\ref{figMeffectiveCompare}.  This figure also shows that as $\mhalf$ increases, the peak increases.

Two very similar observables can also be constructed.  The $b$ effective mass, $\meffb$, is defined by
\begin{equation}
\meffb = p_{T}^{\mathrm{jet1}(b)} + p_{T}^{\mathrm{jet2}} + p_{T}^{\mathrm{jet3}} + p_{T}^{\mathrm{jet4}} + \met,
\label{eqMeffectiveb}
\end{equation}
and the $2b$ effective mass, $\mefftwob$, is similarly defined by
\begin{equation}
\mefftwob = p_{T}^{\mathrm{jet1}(b)} + p_{T}^{\mathrm{jet2}(b)} + p_{T}^{\mathrm{jet3}} + p_{T}^{\mathrm{jet4}} + \met.
\label{eqMeffectivetwob}
\end{equation}
Here there are no restrictions on the non-leading jets; they can be either $b$-tagged or not.  By including the leading $b$ jets, which are primarily decay products of the superpartners to the third generation quarks, we include information about the parameters $\azero$ and $\tanb$.

\begin{figure}[t]
\centering
\includegraphics[width=.70\textwidth]{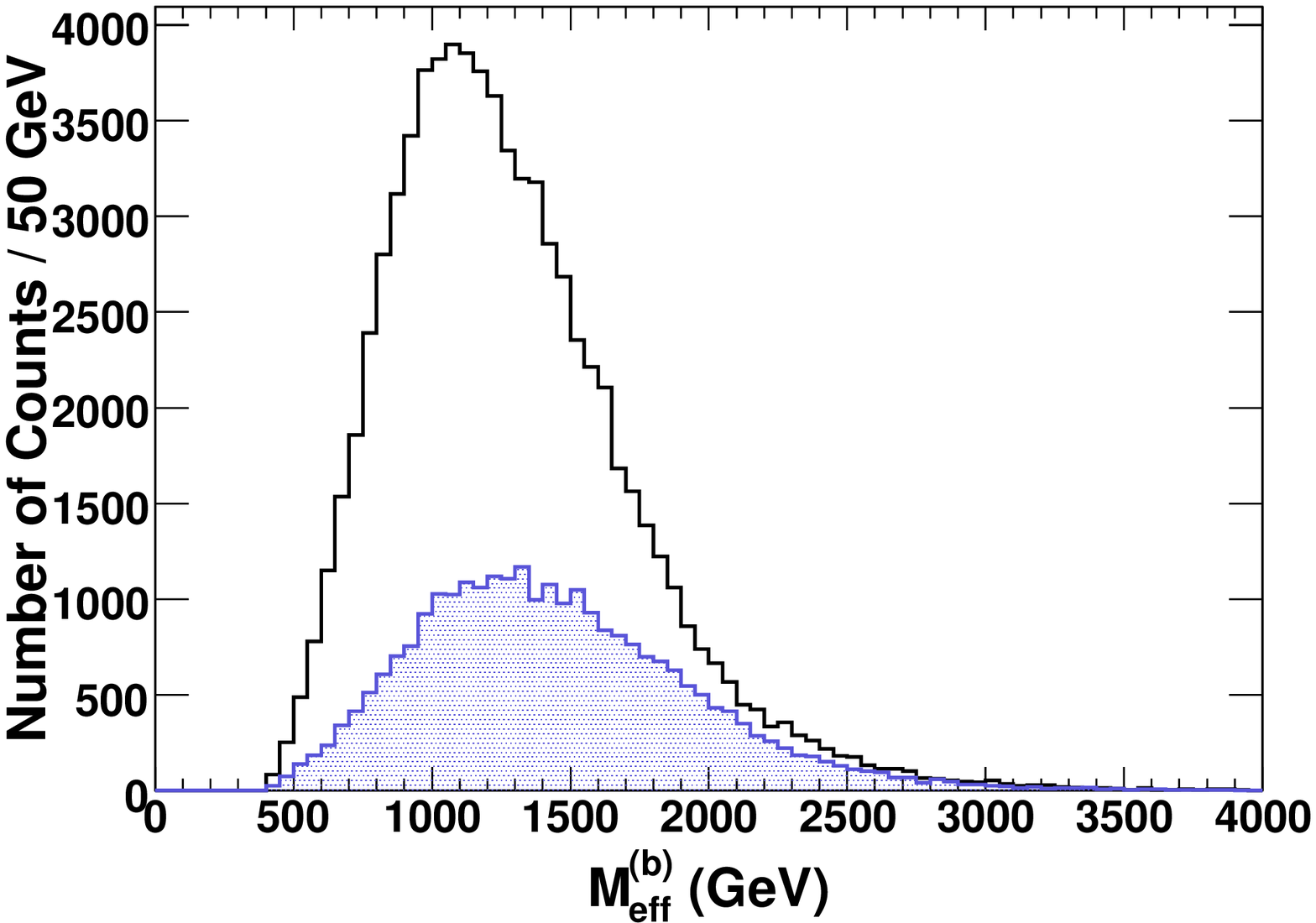}
\caption{Same as Fig.~\ref{figMeffectiveCompare}, except that this is the $b$ effective mass distribution and that the peak fits are not shown.}
\label{figMeffectivebCompare}
\end{figure}

\begin{figure}[t]
\centering
\includegraphics[width=.70\textwidth]{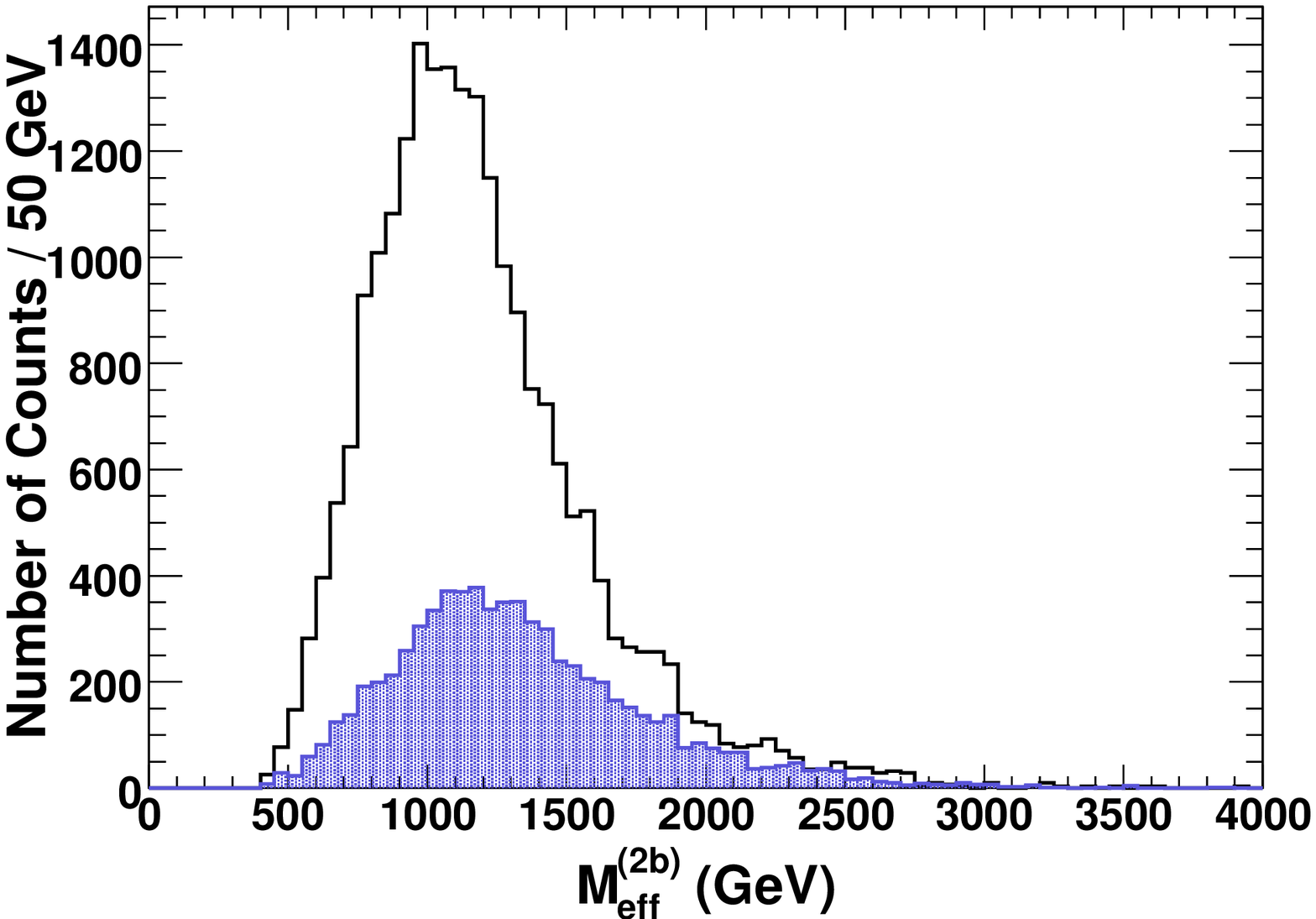}
\caption{Same as Fig.~\ref{figMeffectivebCompare}, except that this is the $2b$ effective mass distribution.}
\label{figMeffective2bCompare}
\end{figure}

To construct these distributions, we use the very same cuts as we used for $\meff$, with some exceptions.  For the $\meffb$ distribution, the leading $\pt$ jet must be tagged as a $b$ jet, otherwise we reject the event.  For the $\mefftwob$ distribution, the two leading $\pt$ jets must both be tagged as $b$ jets.  Again, the non-leading jets in $\meffb$ and $\mefftwob$ can be either $b$-tagged or not.  We also use the same fitting algorithm for these distributions as we have used for $\meff$.  Sample $\meffb$ and $\mefftwob$ effective mass distributions are shown in Figs.~\ref{figMeffectivebCompare} and \ref{figMeffective2bCompare}.
% Figure 10
%
% Figure 11
% Figure 12

Now that we have all the observables we need, we can determine the four mSUGRA parameters in any region.  We will now describe examples of this method for a Higgs boson dominated region, a $\stau$ dominated region, and a $Z$ boson dominated region.

\begin{figure}
\centering
\includegraphics[width=.45\textwidth]{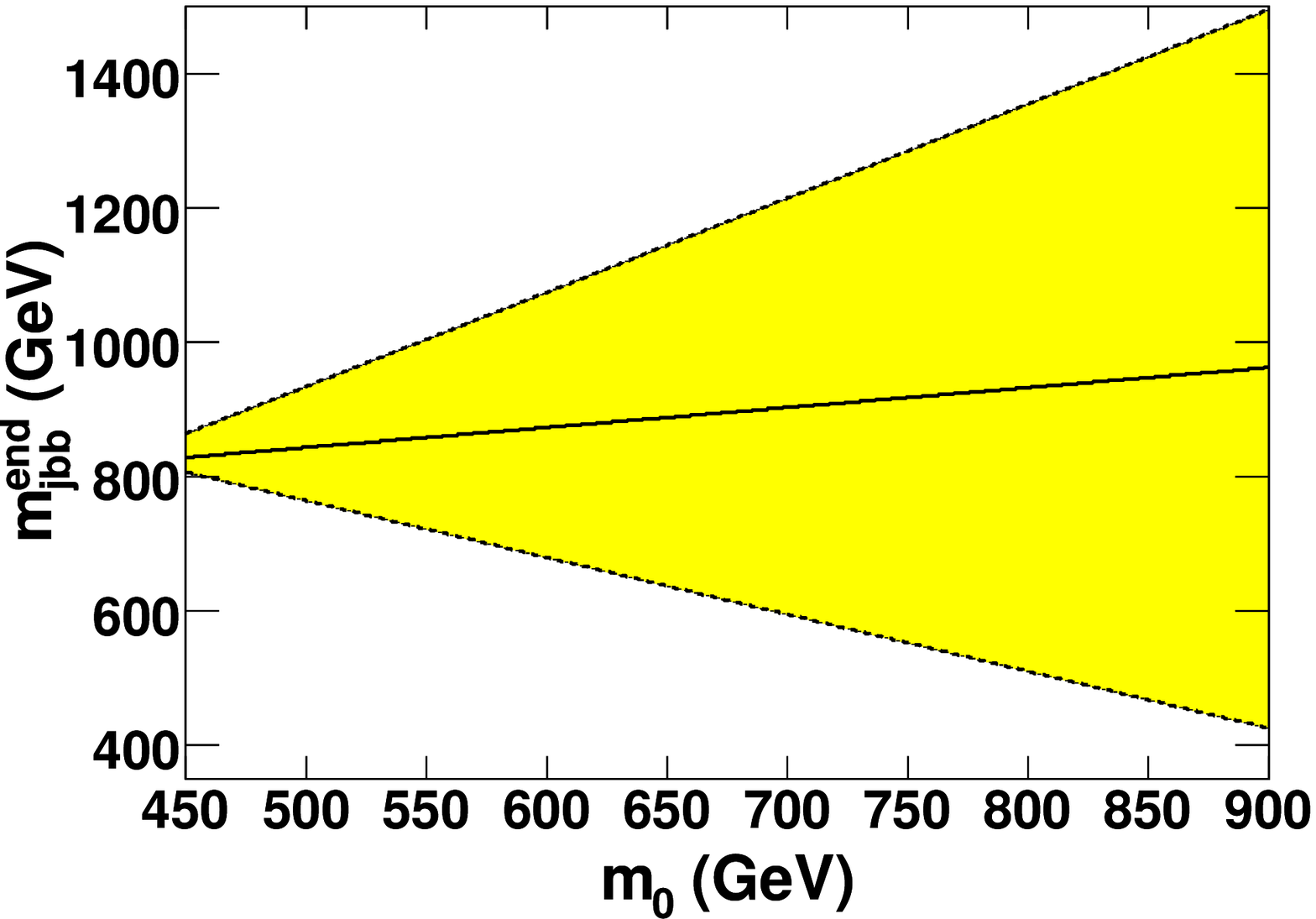}
\hspace{1cm}
\includegraphics[width=.45\textwidth]{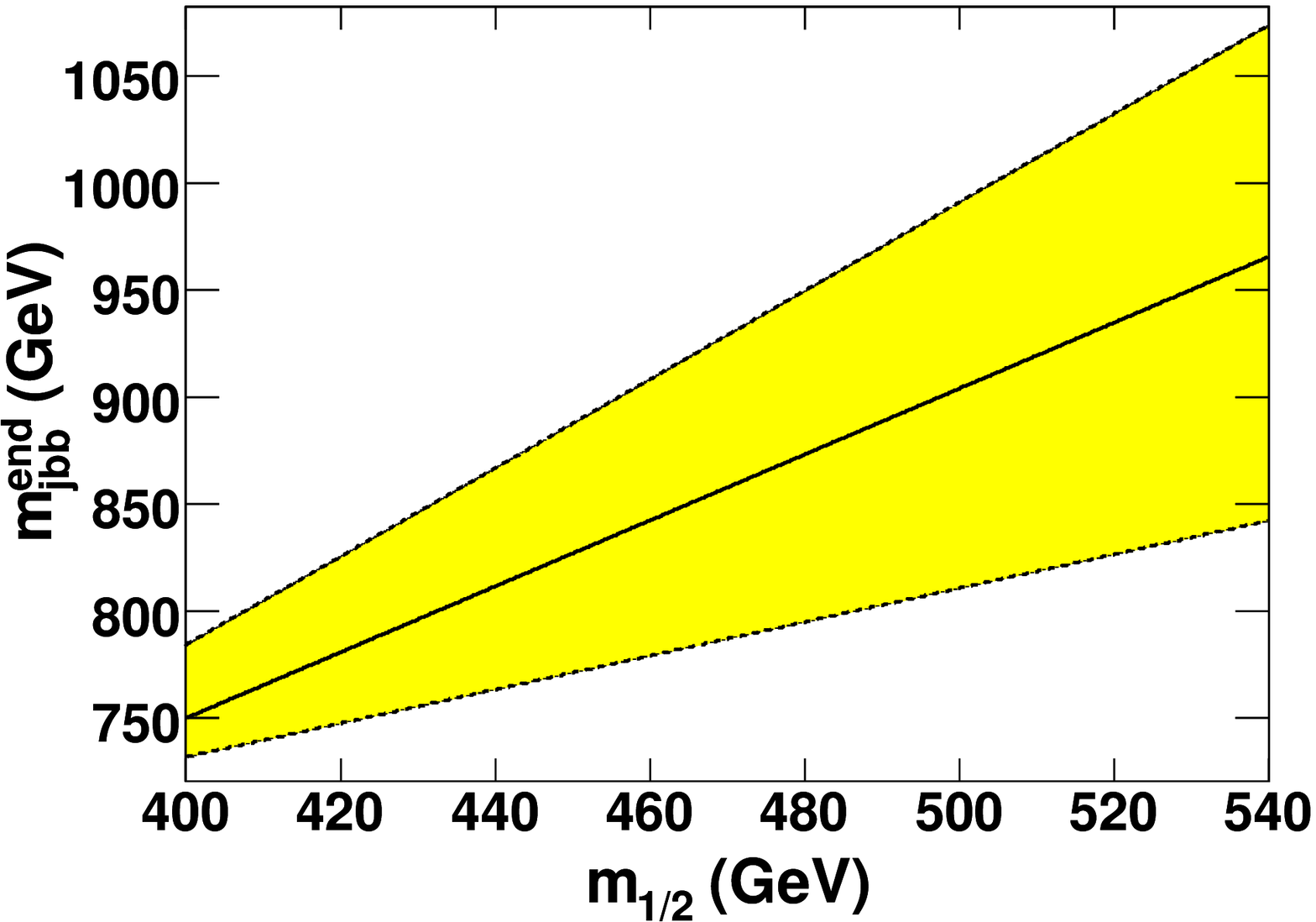}
\caption{The left plot shows the change in $M_{bbj}^{\mathrm{2nd},\mathrm{end}}$ under variations of $\mzero$ ($\mhalf = 440~\gev$) within the Higgs dominant region of parameter space.  The right plot shows the same for variations in $\mhalf$ ($\mzero = 471~\gev$).  Combining the functions plotted results in the functional form $M_{bbj}^{\mathrm{2nd},\mathrm{end}} = f_{4}(\mzero, \mhalf)$.  The $1\sigma$ uncertainty bands (dashed lines) represent $500~\invfb$ of data.}
\label{figHiggsPlusJetFuncForm}
\end{figure}

\begin{figure}
%\centering
\includegraphics[width=.45\textwidth]{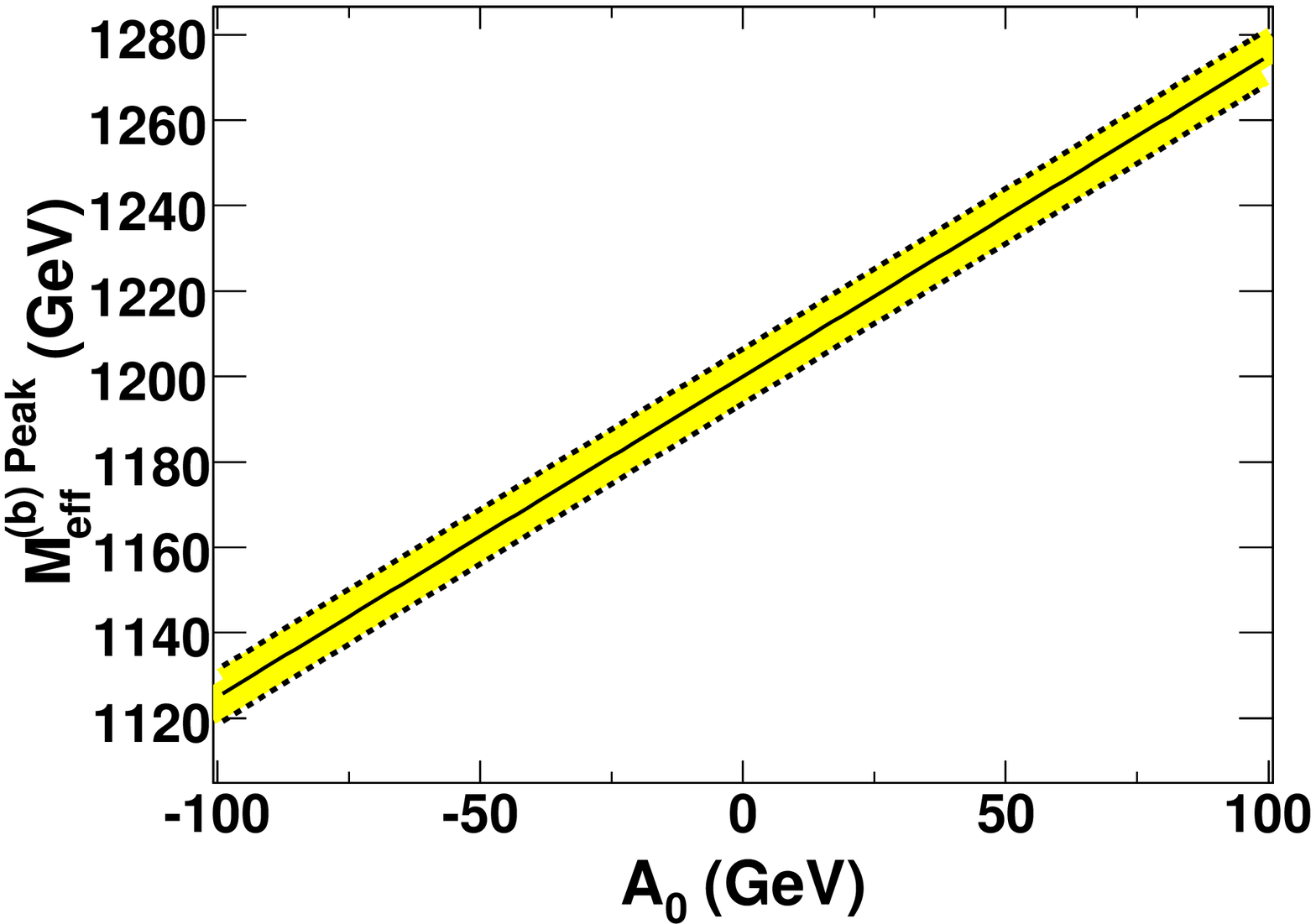}
\hspace{1cm}
\includegraphics[width=.45\textwidth]{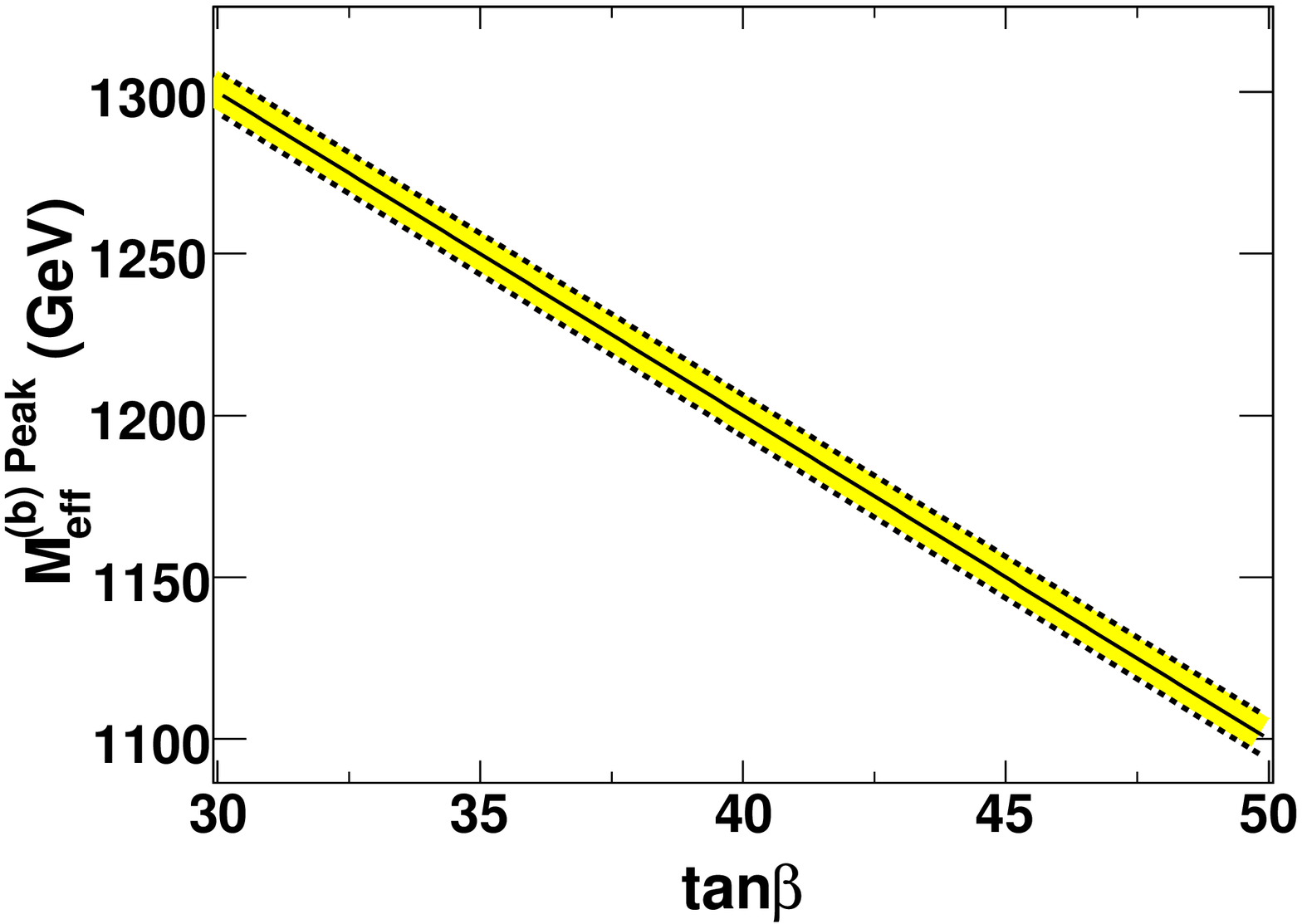}
\caption{The left plot shows the change in $\meffbpeak$ under variations of $\azero$ within the Higgs dominant region of parameter space ($\mzero = 471~\gev$, $\mhalf = 440~\gev$).  The right plot shows the same for variations in $\tanb$.  Combining the functions plotted along with variations in $\mzero$ and $\mhalf$ results in the functional form $\meffbpeak = f_{2}(\mzero, \mhalf, \azero, \tanb)$.  The $1\sigma$ uncertainty bands (dashed lines) represent $1000~\invfb$ of data.}
\label{figMeffbFuncForm}
\end{figure}

{\bf Higgs + jets + $\met$}:
For the Higgs dominant region, we use the following four observables to determine our mSUGRA parameters:\begin{itemize}
\item Effective Mass: $\meffpeak = f_{1}(\mzero, \mhalf)$
\item $b$ Effective Mass: $\meffbpeak = f_{2}(\mzero, \mhalf, \azero, \tanb)$
\item $2b$ Effective Mass: $\mefftwobpeak = f_{3}(\mzero, \mhalf, \azero, \tanb)$
\item Higgs plus jet invariant mass: $M_{bbj}^{\mathrm{2nd},\mathrm{end}} = f_{4}(\mzero, \mhalf)$
\end{itemize}  These functional forms are determined by examining how each kinematical observable changes while varying one of the mSUGRA parameters.  Examples of this are shown in Figs.~\ref{figHiggsPlusJetFuncForm} and \ref{figMeffbFuncForm}.
%%%% Figure 13
%%%% Figure 14
%%%%
To determine our mSUGRA parameters, we invert these functional forms into functions of the mSUGRA parameters in terms of the kinematical observables.  Then we can simply plug in the values of the observables into the inverted functions to solve for the mSUGRA parameters.  To get the uncertainties of the mSUGRA determinations, we propagate the uncertainties of the measured  observables through the inverted functions using a Monte Carlo method.

\begin{figure}
%\centering
\includegraphics[width=.45\textwidth]{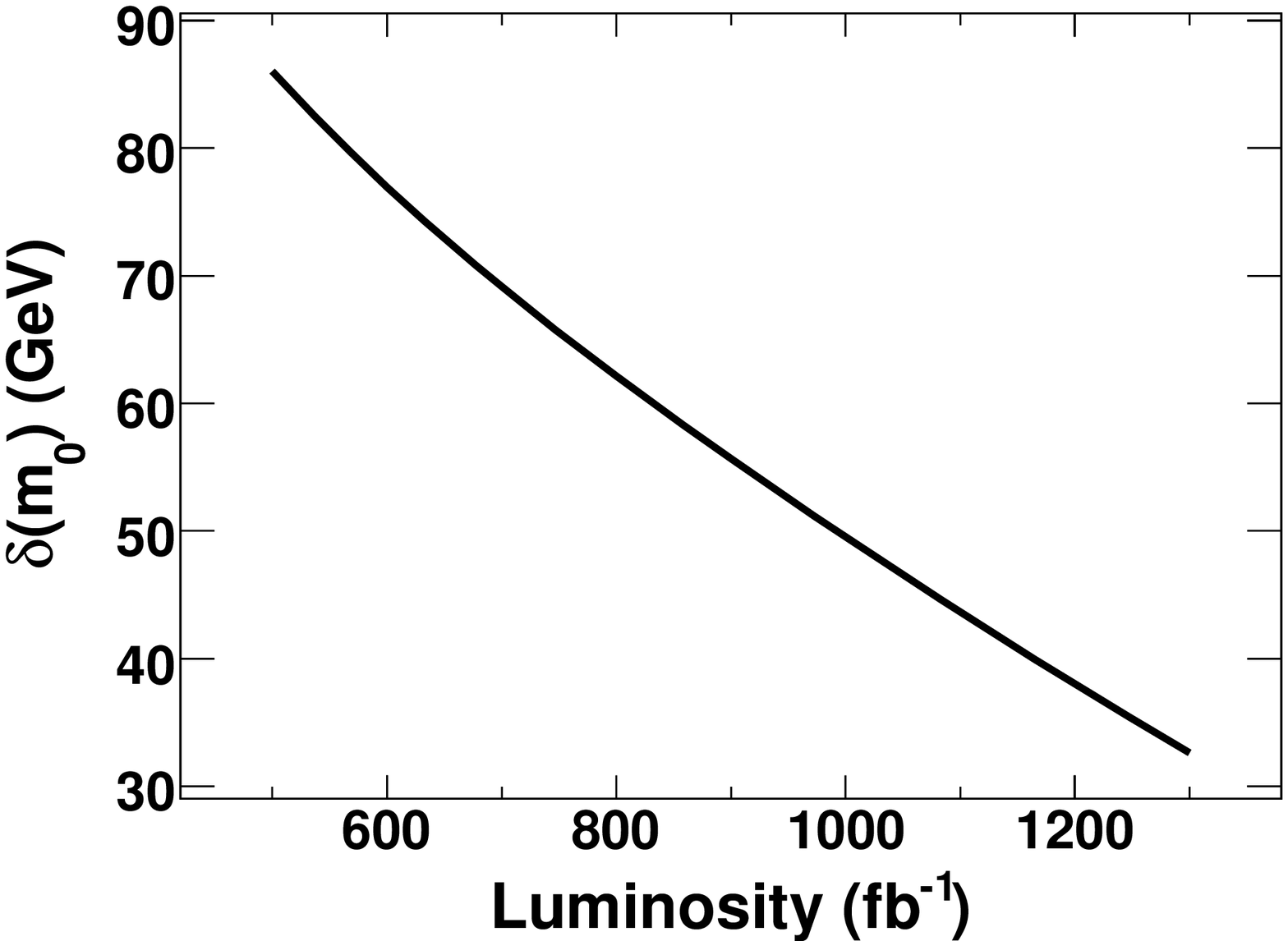}
\hspace{1cm}
\includegraphics[width=.45\textwidth]{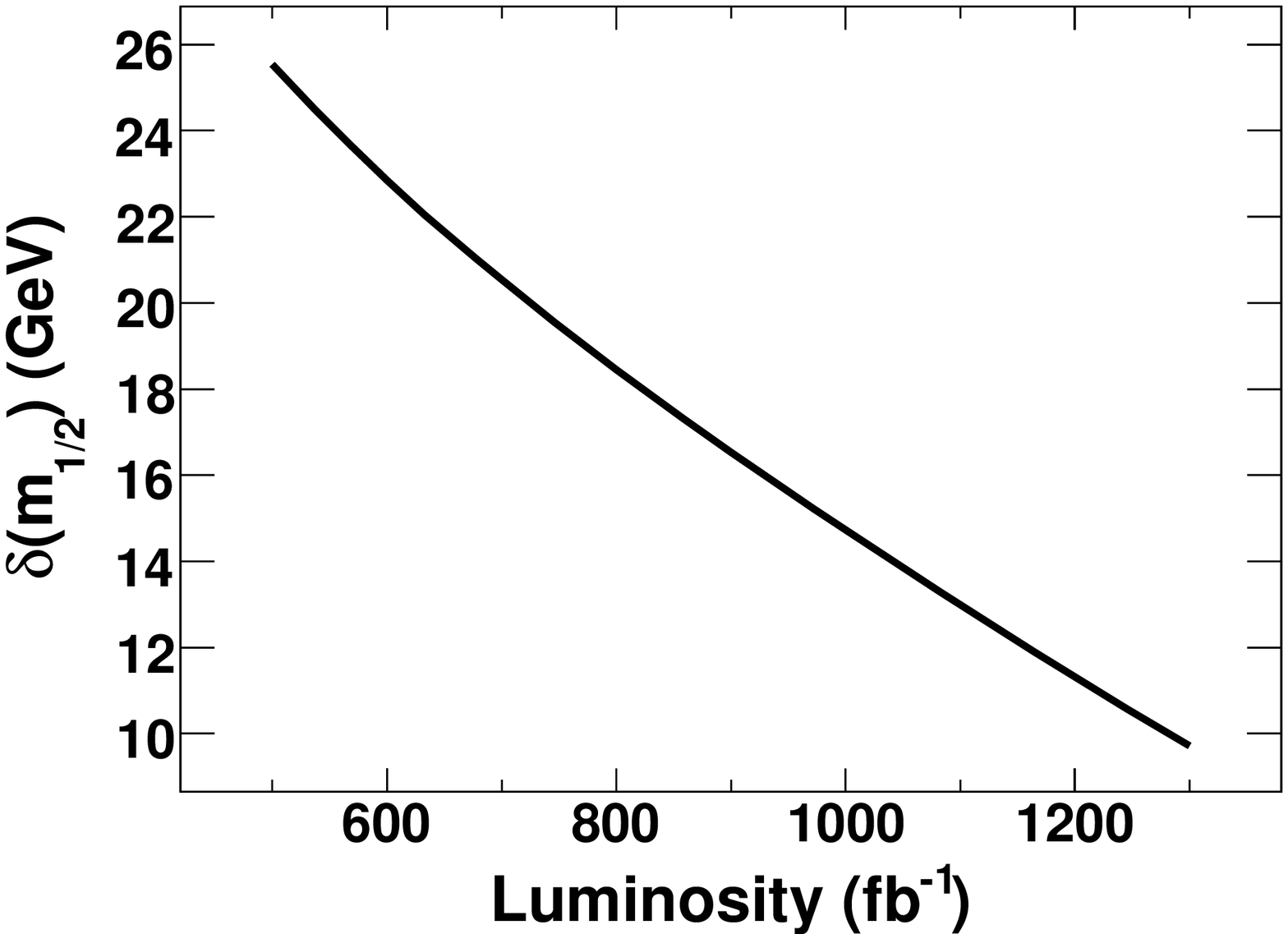}
\caption{The left plot shows the change in the measurement uncertainty of $\mzero$ in the Higgs dominant region of parameter space ($\mzero = 471~\gev$, $\mhalf = 440~\gev$) for different luminosities.  The right plot shows the same for the uncertainty in $\mhalf$.}
\label{figUncVSLum:HiggsM0Mhf}
\end{figure}

\begin{figure}
%\centering
\includegraphics[width=.45\textwidth]{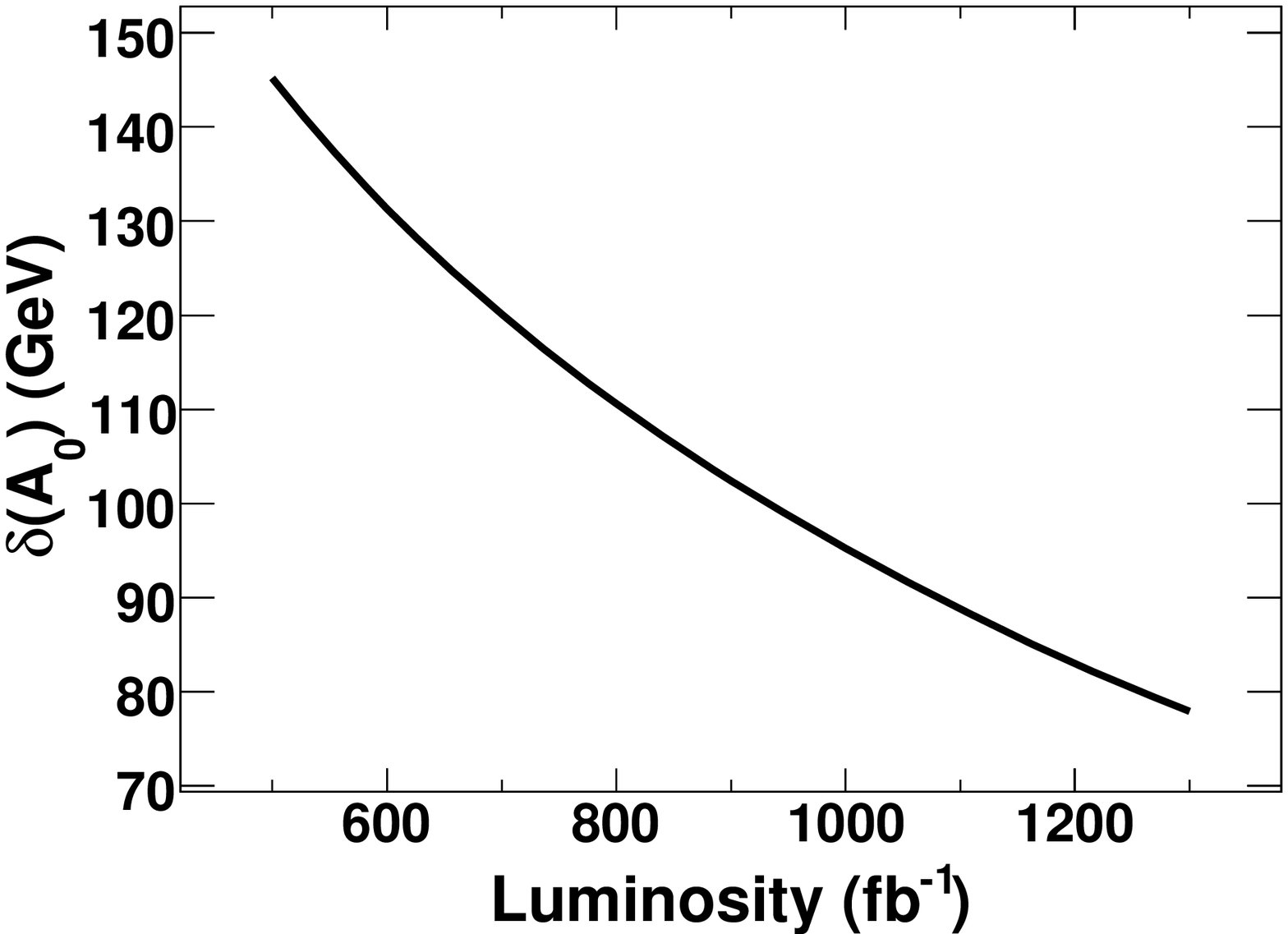}
\hspace{1cm}
\includegraphics[width=.45\textwidth]{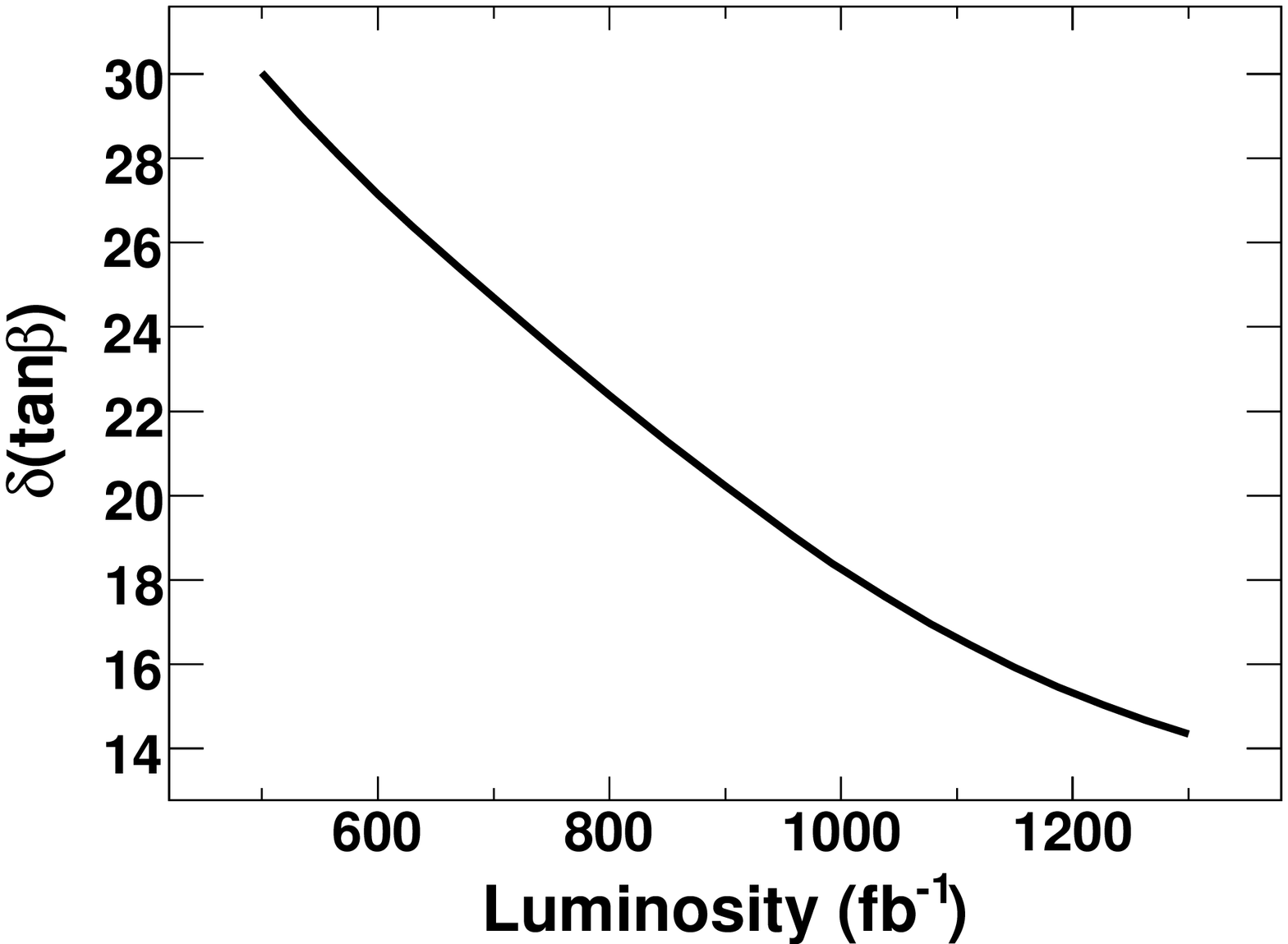}
\caption{Same as Fig.~\ref{figUncVSLum:HiggsM0Mhf} but for $\azero$ and $\tanb$.}
\label{figUncVSLum:HiggsA0tanb}
\end{figure}

\begin{figure}
\centering
\includegraphics[width=.45\textwidth]{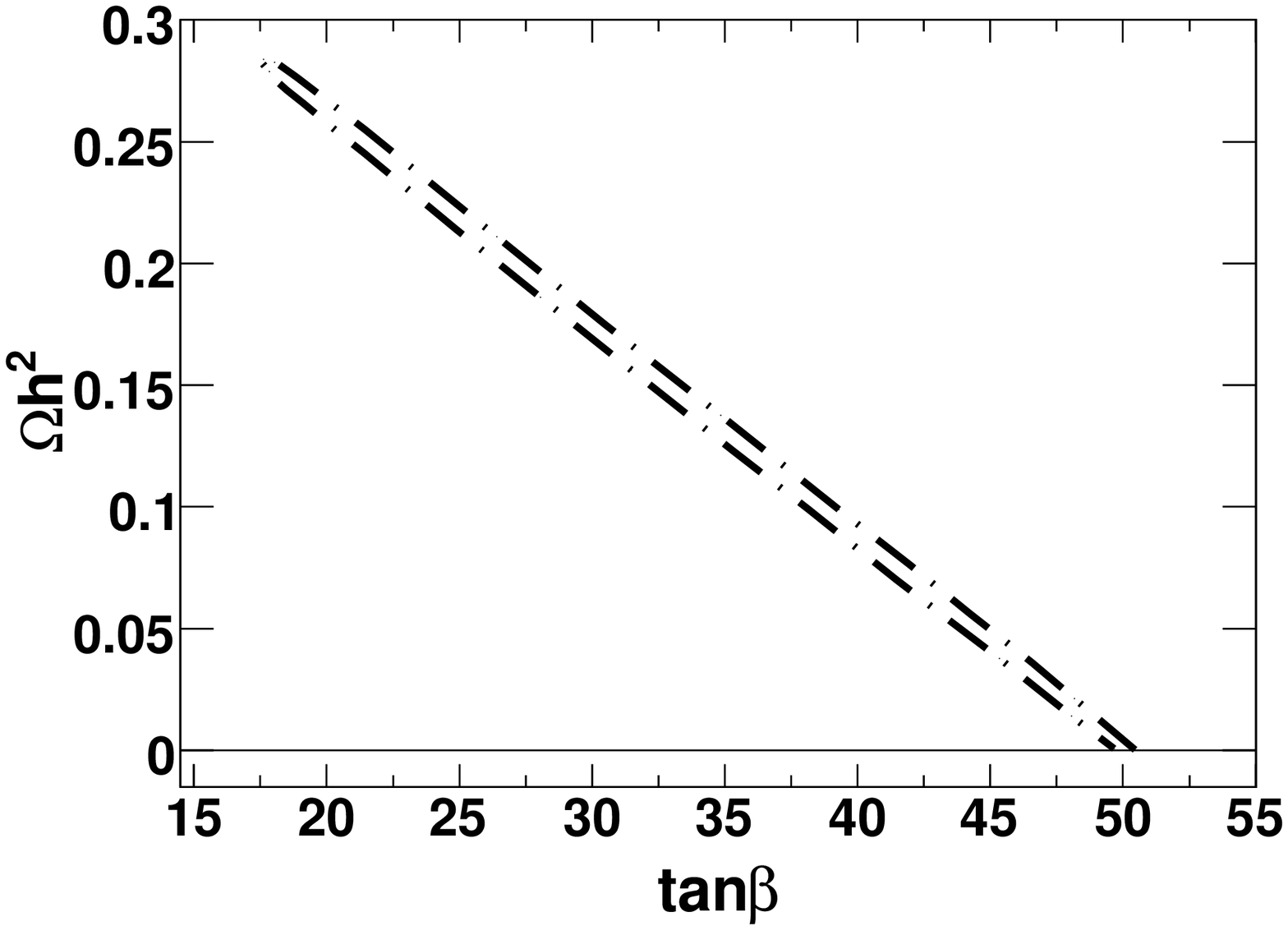}
\hspace{1cm}
\includegraphics[width=.45\textwidth]{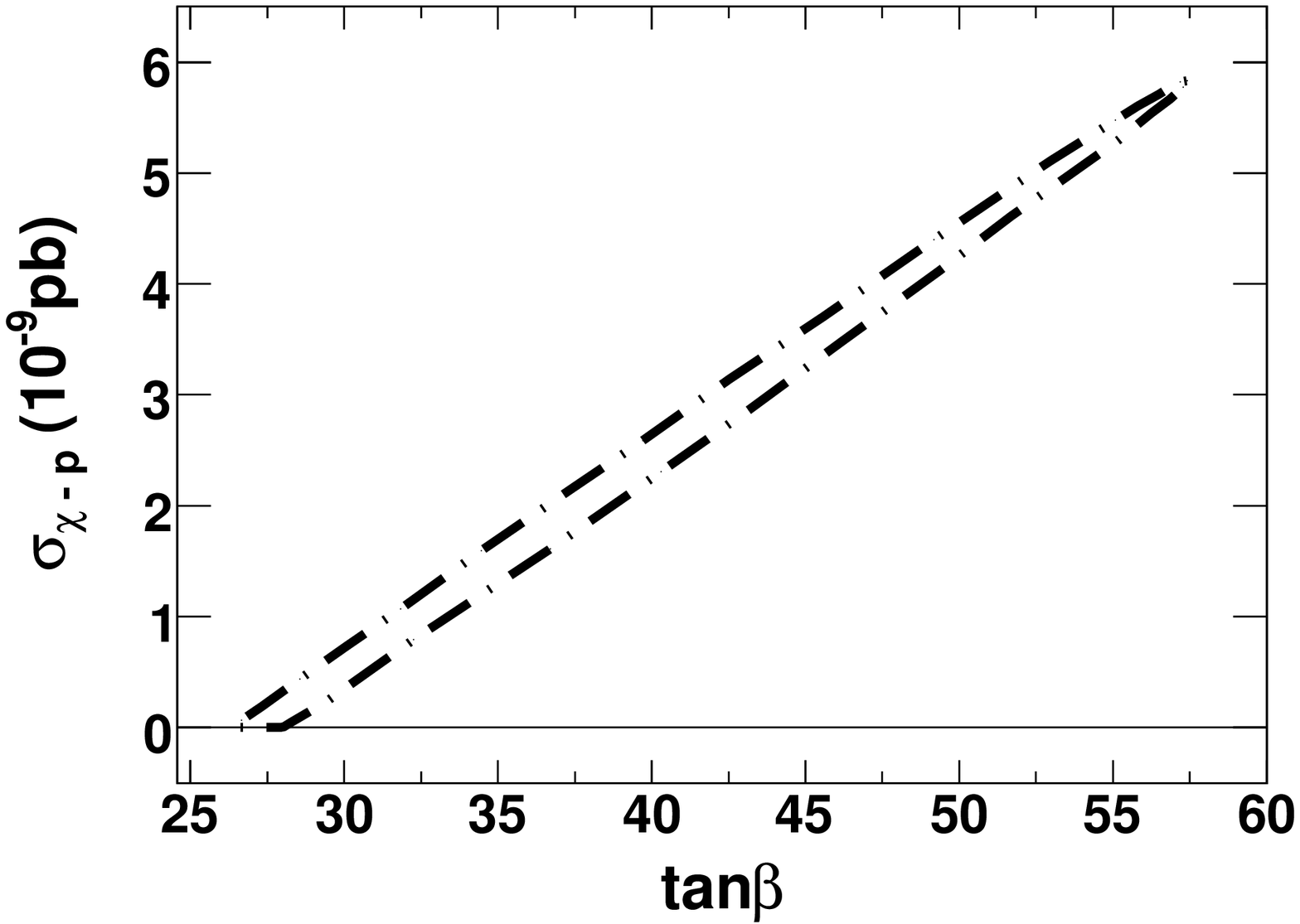}
\caption{The left plot shows the $1\sigma$ uncertainty ellipse on the $\DMrelic$-$\tanb$ plane in the Higgs dominant region of parameter space ($\mzero = 471~\gev$, $\mhalf = 440~\gev$).  The right plot shows the same for the $\pncross$-$\tanb$ plane.  The uncertainty in $\tanb$ is the main source of both the uncertainty of $\DMrelic$ as well as the uncertainty of $\pncross$.  These results are for $1000~\invfb$ of data.}
\label{figErrorEllipse:HiggsRegion}
\end{figure}

We perform a sample analysis for the Higgs region with the following result:  $\mzero = 472\pm50~\gev$, $\mhalf = 440\pm15~\gev$, $\azero = 0\pm95~\gev$, and $\tanb =39\pm17$.  These uncertainties were achieved at $1000~\invfb$.  The relation between the uncertainties and the luminosity is shown for these parameters in Figs.~\ref{figUncVSLum:HiggsM0Mhf} and \ref{figUncVSLum:HiggsA0tanb}.  Using these results, we can also calculate the neutralino relic density and proton-neutralino cross section.  The result for $1000~\invfb$ is:  $\DMrelic = 0.10\pm0.15$ and $\pncross = (1.9\pm3.7)\times10^{-9}~\mathrm{pb}$.  The uncertainty ellipses on the $\DMrelic$-$\tanb$ and $\pncross$-$\tanb$ planes are shown in Fig.~\ref{figErrorEllipse:HiggsRegion}.  Since the uncertainties in each of these values are larger than 100\%, the uncertainty ellipses get pushed into negative (unphysical) values of $\DMrelic$ and $\pncross$.  As such, we have cut these ellipses off at the $x$-axes in Fig.~\r
 ef{figErrorEllipse:HiggsRegion}.
%%%% Figure 15
%%%% Figure 16
%%%%
%%%% Figure 17

%%%%

\begin{figure}
\centering
\includegraphics[width=.70\textwidth]{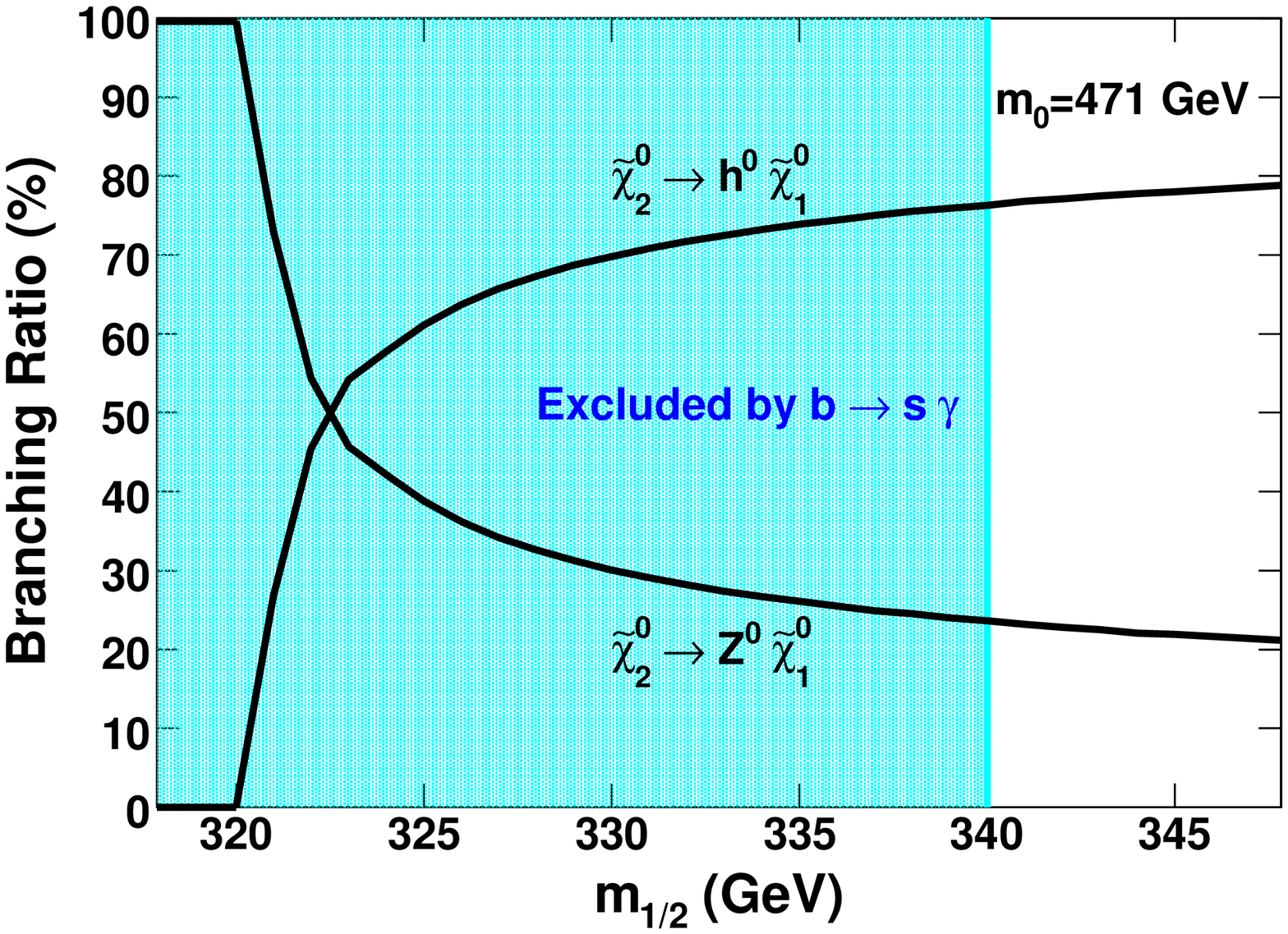}
\caption{The branching ratios for $\schitwozero\rightarrow h^{0} \schionezero$ and $\schitwozero\rightarrow Z \schionezero$ as a function of $\mhalf$ for $\mzero = 471~\gev$, $\azero = 0$, and $\tanb = 40$.  Also shown is the $b \rightarrow s \gamma$ exclusion region (cyan filled region) from Fig.\ref{figParameterSpace}.}
\label{figZspaceBR}
\end{figure}

{\bf $Z$ + jets + $\met$}:
The analysis technique in the $Z$ dominant region is just the same as the Higgs region if we replace the Higgs plus jet invariant mass with the $Z$ plus jet invariant mass.  The endpoint of the latter can be measured with better precision.  This is due to both an increase in production cross section and the ease of reconstructing $Z$ bosons from lepton pairs.  This results in a more precise determination of $\DMrelic$ and $\pncross$ comparable to that of the $\stau$ dominant region shown below.  However, we suffer from small $\Br\left(Z \rightarrow ll \right)$ values.  For a useful measurement, we need $\Br\left(\schitwozero\rightarrow Z \schionezero\right) \gtrsim 50\%$.  However, such a region does not exist outside of the $b \rightarrow s \gamma$ bound, as shown in Fig. \ref{figZspaceBR}. For lower $\tanb$, the same conclusion holds since we get constraints on the smaller values of $\mhalf$ due to Higgs mass. Therefore, we do not go into detailed analysis of the determinat
 ion of model parameters in the $Z$ + jets + $\met$ region.  However, one can use the observables of the Higgs + jets + $\met$ region, e.g., $\meff$, $\meffb$, and $\mefftwob$, to reconstruct the model parameters.

%%%% Figure 18

%%%%

\begin{figure}
%\centering
\includegraphics[width=.45\textwidth]{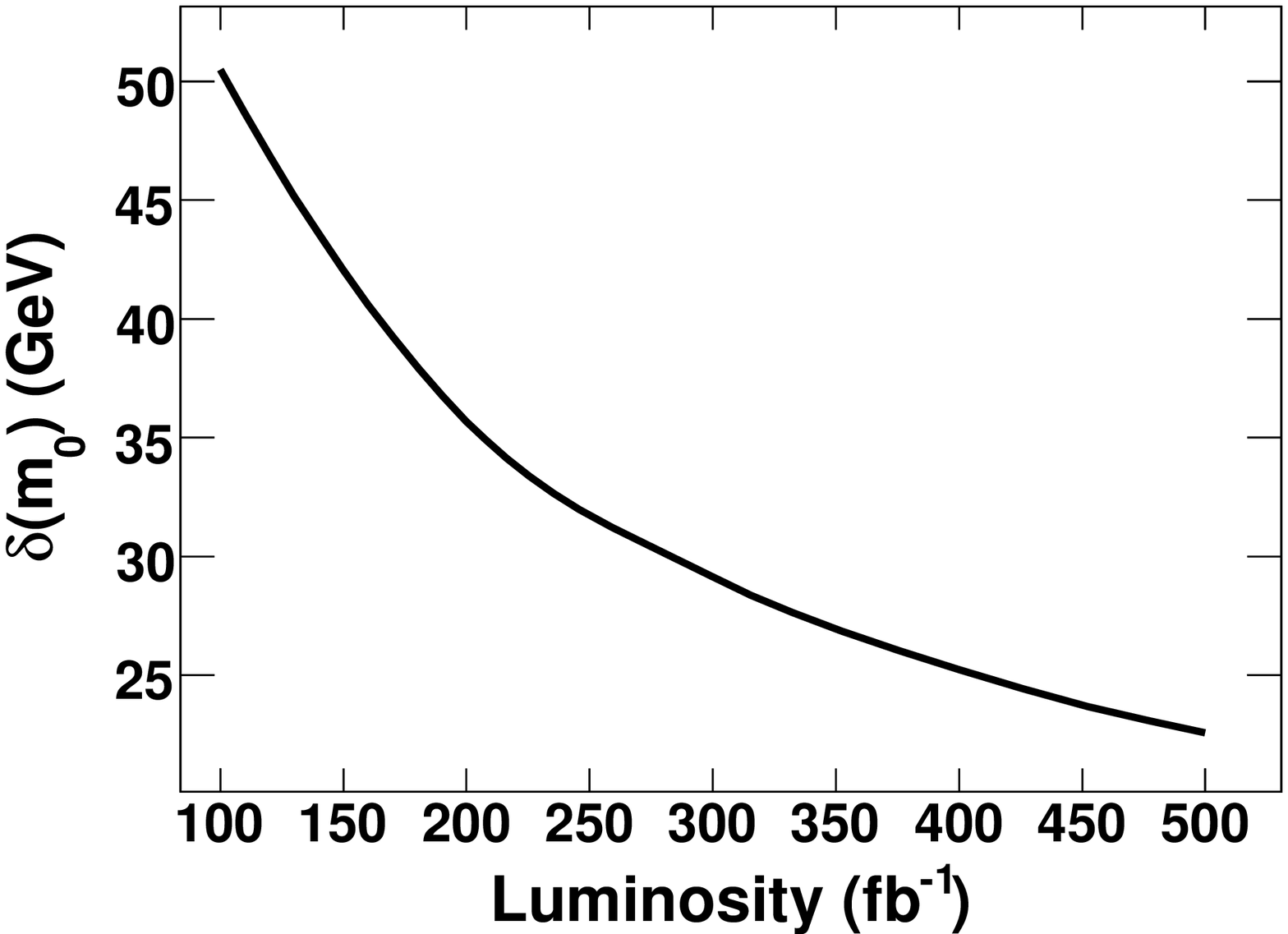}
\hspace{1cm}
\includegraphics[width=.45\textwidth]{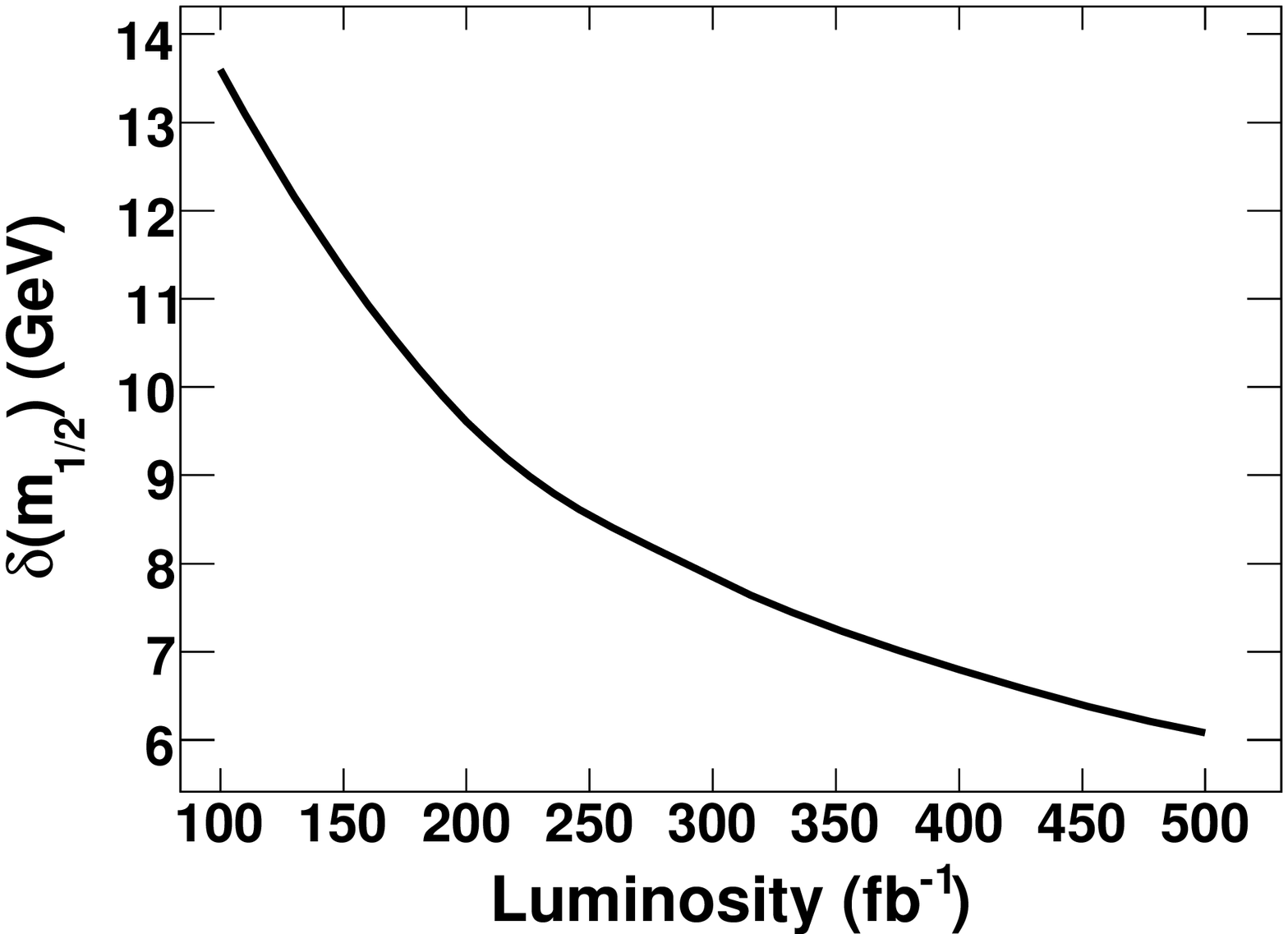}
\caption{Same as Fig.~\ref{figUncVSLum:HiggsM0Mhf} except within the $\stau$ dominant region of parameter space ($\mzero = 440~\gev$, $\mhalf = 600~\gev$).}
\label{figUncVSLum:StauM0Mhf}
\end{figure}

\begin{figure}
%\centering
\includegraphics[width=.45\textwidth]{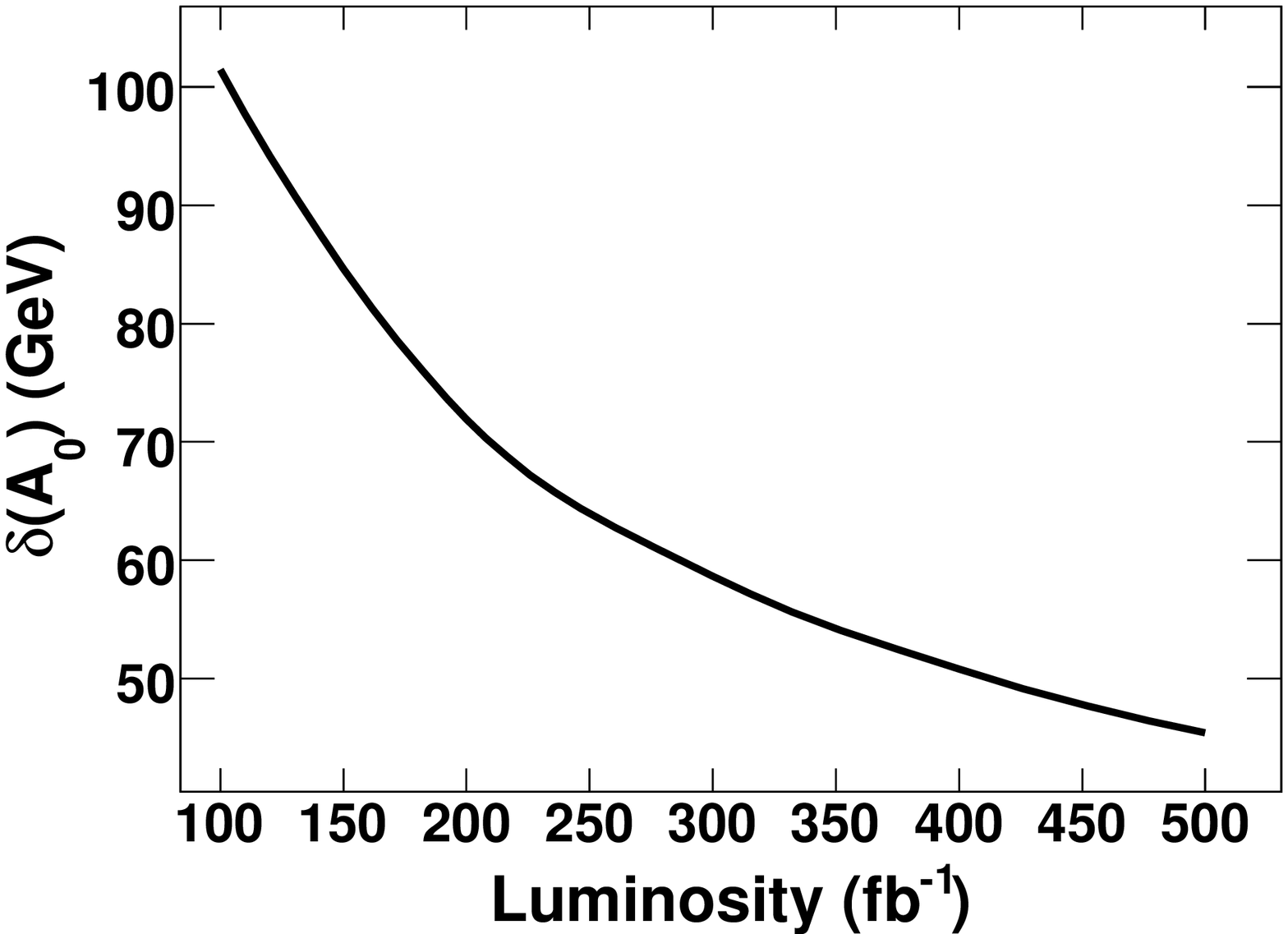}
\hspace{1cm}
\includegraphics[width=.45\textwidth]{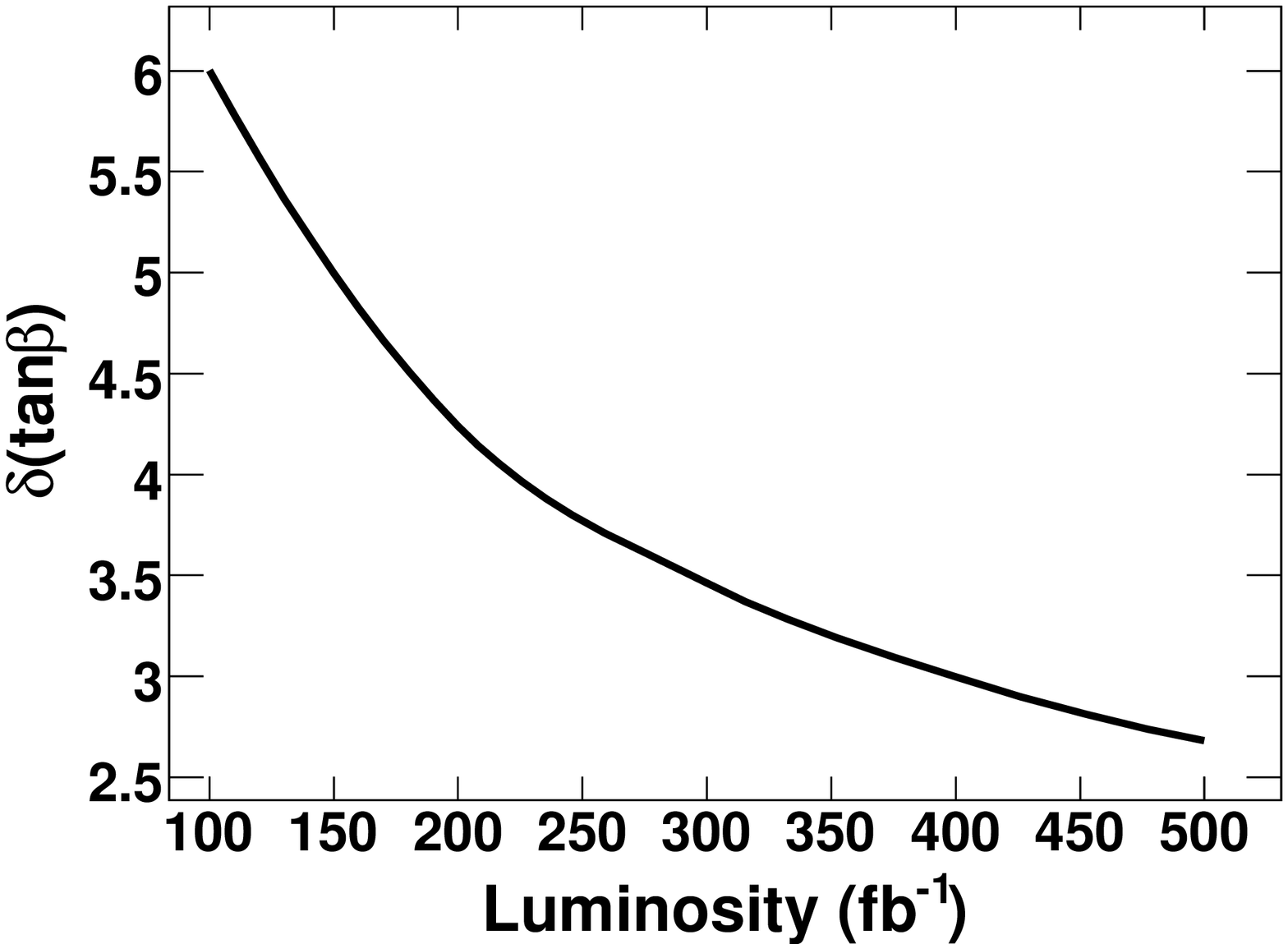}
\caption{Same as Fig.~\ref{figUncVSLum:HiggsM0Mhf} except for $\azero$ and $\tanb$ within the $\stau$ dominant region of parameter space ($\mzero = 440~\gev$, $\mhalf = 600~\gev$).}
\label{figUncVSLum:StauA0tanb}
\end{figure}

\begin{figure}
\centering
\includegraphics[width=.45\textwidth]{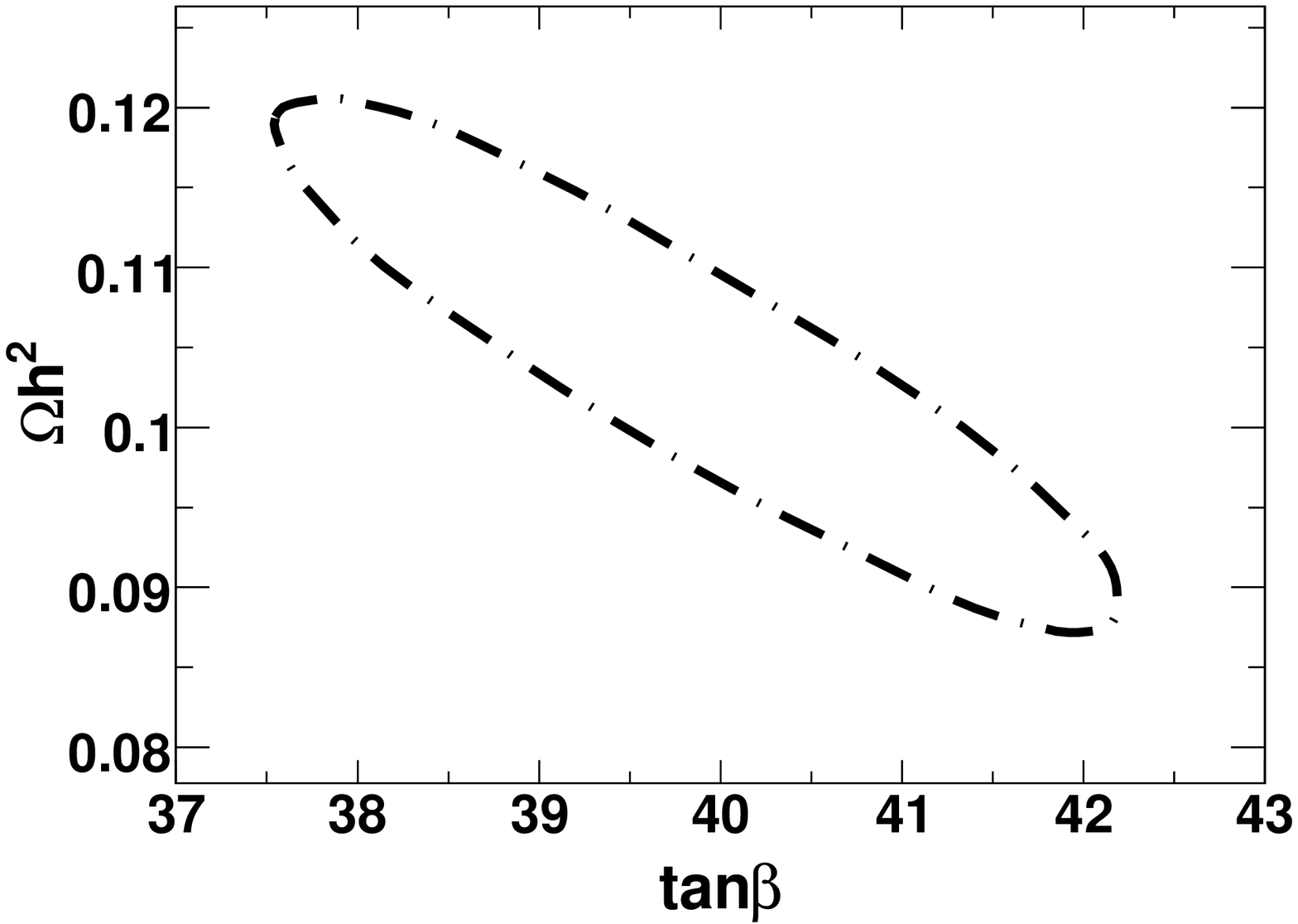}
\hspace{1cm}
\includegraphics[width=.45\textwidth]{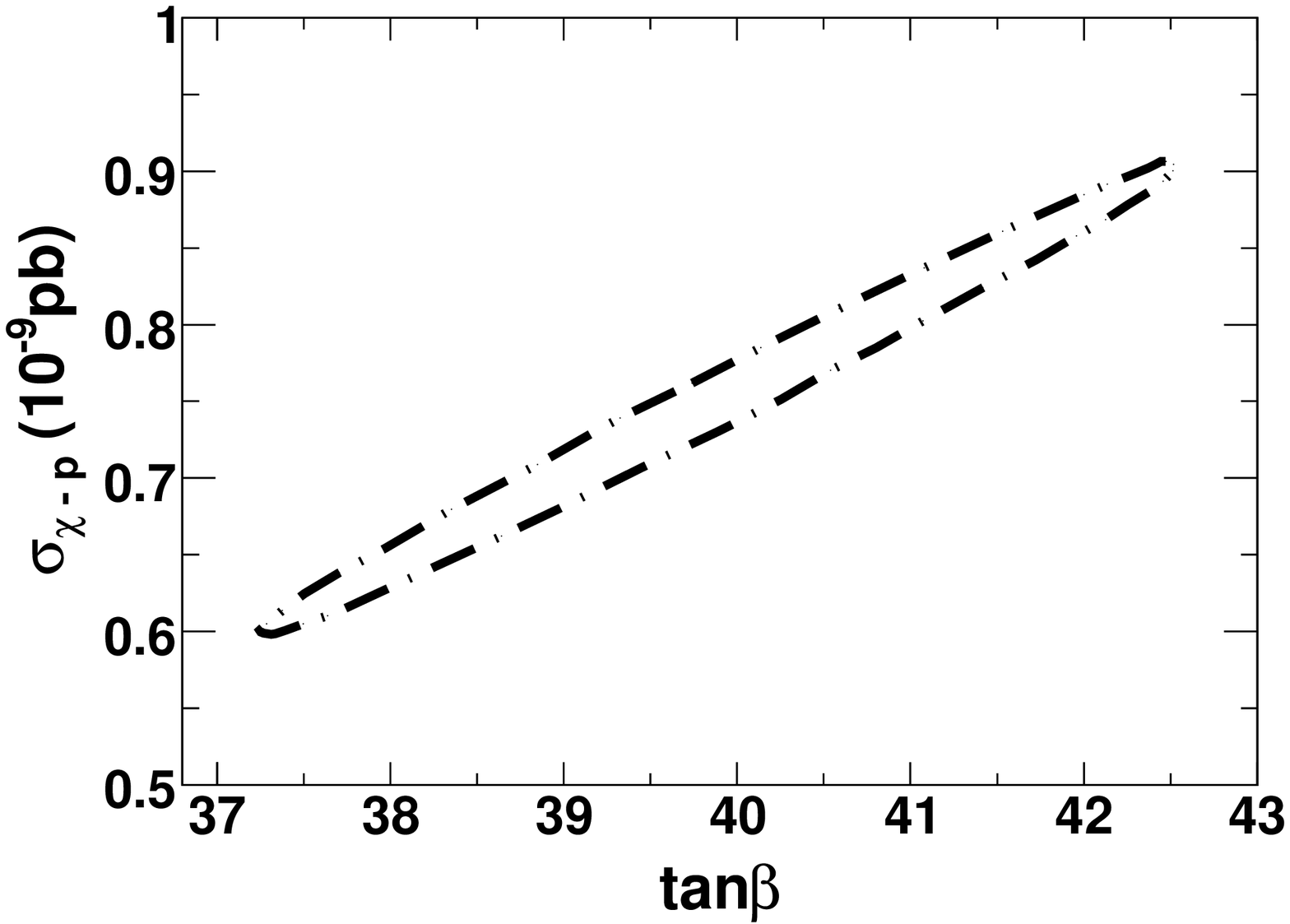}
\caption{Same as Fig.~\ref{figErrorEllipse:HiggsRegion} except within the $\stau$ dominant region of parameter space ($\mzero = 440~\gev$, $\mhalf = 600~\gev$).  These results are for $500~\invfb$ of data.}
\label{figErrorEllipse:StauRegion}
\end{figure}

{\bf 2 $\tau$ + jets + $\met$}:
For the $\stau$ dominant region, we use the following four observables to determine our mSUGRA parameters:\begin{itemize}
\item Effective Mass: $\meffpeak = f_{1}(\mzero, \mhalf)$
\item $b$ Effective Mass: $\meffbpeak = f_{2}(\mzero, \mhalf, \azero, \tanb)$
\item Ditau Invariant Mass: $M_{\tau\tau}^{\mathrm{peak}} = f_{3}(\mzero, \mhalf, \azero, \tanb)$
\item Ditau plus jet invariant mass: $M_{j\tau\tau}^{\mathrm{peak}} = f_{4}(\mzero, \mhalf)$
\end{itemize}
We again perform an inversion to determine our mSUGRA parameters, as well as propagate the uncertainties in the same way as in the Higgs region.  Our sample analysis for this region yields: $\mzero = 440\pm23~\gev$, $\mhalf = 599.9\pm6.1~\gev$, $\azero = 0\pm45~\gev$, $\tanb = 40.0\pm2.7$, $\DMrelic = 0.103\pm0.019$, and $\pncross = (7.6\pm1.6)\times10^{-10}~\mathrm{pb}$.  These uncertainties were achieved at $500~\invfb$. The parameter $\tan\beta$ is determined with much higher accuracy since we can use observables involving staus; the staus are very sensitive to $\tan\beta$.  Again we show the relation between the uncertainties and the luminosity for this result in Figs.~\ref{figUncVSLum:StauM0Mhf} and \ref{figUncVSLum:StauA0tanb}.  We also again show the uncertainty ellipses on the $\DMrelic$-$\tanb$ and $\pncross$-$\tanb$ planes in Fig.~\ref{figErrorEllipse:StauRegion}. Since $\tan\beta$ is determined with better accuracy compared to the Higgs dominant region, the relic d
 ensity and proton-neutralino cross section are also determined with a better accuracy.

%%%% Figure 19

%%%%

%%%% Figure 20

%%%% Figure 21

%\newpage

\section{Conclusions and Discussion}

In this paper, we have studied the mSUGRA final states at the LHC which are motivated by supercritical string cosmology (SSC).  In the SSC case, the time dependent dilaton not only contributes to the dark energy but also to the Boltzman equation which determines the dark matter content of the universe.  Consequently the dark matter profile in this model is different compared to the standard cosmology.  We found that the dark matter allowed region has larger values of $\mzero$ compared to the standard cosmology case.  Thus, the final states in the SSC scenario are different from those of the standard cosmology.  For example, in the case of standard cosmology for smaller values of $m_0$ (also allowed by the $g_{\mu}-2$ constraint), we have low energy taus in the final state due to the
proximity of the stau to the neutralino mass in the stau-neutralino coannihilation region. On the other hand, in the SSC case the final states contain $Z$ bosons, Higgs bosons or high energy taus. In fact these final states dominate in most of the allowed SUGRA parameter space. Therefore, by analyzing the parameter space of the SSC model we actually investigate most regions of the SUGRA parameter space at the LHC.  We analyzed the signals involving Higgs + jets + $\met$, $Z$ + jets + $\met$, and $2\tau$ + jets + $\met$ and constructed observables such as the endpoints of invariant mass distributions $M_{bbj}$, $M_{Zj}$, and $M_{j\tau\tau}$ and the peak point of $M_{\tau\tau}$.

In order to determine all parameters of the mSUGRA model we needed additional obsevables such as the peak positions of the effective masses $\meffb$ and $\mefftwob$. These observables are used for determining $\tanb$. We found that $\mzero$, $\mhalf$,  and $\tanb$ can be determined with 11\%, 3\% and 44\% accuracies respectively in the Higgs boson dominated final states region for $1000~\invfb$ of data. The $Z$ boson dominated final state region is mostly ruled out by other experimental data.  However, the technique used to analyze the $Z$ boson dominated region is nearly identical to that of the Higgs boson dominated region.  In the stau dominated region, $\mzero$, $\mhalf$,  and $\tanb$ can be determined with  5\%, 1\% and 7\% accuracies  respectively for $500~\invfb$ of data. The accuracy of determining $\tanb$ is improved in the tau dominated final state region since we use observables involving the staus which are very sensitive to the variation of $\tanb$. Once all the
 parameters are known, the dark matter content can be determined in all these cases. In the Higgs dominant case, the accuracy of determining the dark matter content is 150\% for $1000~\invfb$ of data.  In contrast, the accuracy of relic density in the stau dominated region is 18\% for $500~\invfb$ of data, which is much better due to a higher accuracy of $\tanb$ determination. These techniques can be applied in the case of nonuniversal models as well, where we will need more observables to determine the model parameters.

When the LHC will be operating, we will also have results from the dark matter direct detection experiment. We found that the cross section for these experiments can be predicted from the LHC measurements with an accuracy of 195\% for $1000~\invfb$ of data in the Higgs boson dominated region and 21\% for $500~\invfb$ of data in the stau dominated region. This cross section however includes uncertainity due to the form factors.

As a remark, our phenomenological study assumed several key detector performances of the present ATLAS and CMS detectors, such as b-tagging and tau identification efficiencies. We find
that one needs 500-1000\ \invfb\ of data. The regime of such high luminosity can be realized with the LHC's luminosity upgrade as well as the upgrade of both ALTAS and CMS detectors.
Thus, our results are just a guideline for the physics case if the performance of both upgraded detectors is the same even at such high luminosity operation of the luminosity-upgraded LHC.

In this analysis we examined the overdense region of the mSUGRA model
since the underlying cosmological theory converts the overdense region
into a region with  correct relic abundance. This analysis holds for
any cosmological model with similar features.

Before closing we repeat some cautionary remarks regarding the microscopic model dependence of such studies~\cite{mavromatoslhc}. As already mentioned in the introduction, the low-energy limit of string theory is incredibly non-unique, as it depends on the complicated details of compactification and SUSY breaking procedures. Various models lead to different predictions, and some of them may lead to completely different phenomenology as far as dark matter studies are concerned. For instance, there are heterotic string models entailing non-thermal dark matter~\cite{hetero}, whose detection requires totally different techniques from the ones employed here.

Nevertheless, there are string models which can be analyzed rather generically within the methods outlined in this work, in the sense that the observables discussed in this analysis can also be used to extract information on dark matter in such string-inspired models as well. For instance, the moduli-dominated sector of the heterotic (orbifold-compactified) class of models examined in \cite{hetero}  has
five parameters  the gravitino mass $m_{3/2}$, the vacuum expectation value of the real part of the (uniform) Kahler modulus $\langle t + {\bar t}\rangle$, the modular weights of the Pauli-Villars regulators parameterized by p, the value of the Green-Schwarz coefficient  $\delta_{GS}$ and  $\tan\beta$. The parameters of this model can be determined in the same spirit as shown in the paper and thereby the dark matter density can also be determined in the way we have described in this work. The same procedure can be applied to the dilaton dominated models described in the same reference.  Depending on the model which is used, we may need to construct more observables to determine all the model parameters.

%The real problem, as far as dark-matter phenomenology is concerned, is for those models which have
% more parameters than observables at the LHC. Affaire \'a suivre ...

\section*{Acknowledgments}
This work is supported in part by the DOE grant
DE-FG02-95ER40917 and NSF grant DMS 0216275.  The work of A.G. is supported by DOEd GAANN. The work of A.B.L. and N.E.M is supported in part by the European Union through the FP6 Marie-Curie Research and Training Network, Universenet (MRTN-CT-2006-035863), and that of A.B.L. also in part by the European Union Research and Training Network MRTN-CT-2004-503369.


\begin{thebibliography}{99}
\bibitem{uncertainties}
All uncertainties are statistical only.
The systematic uncertainties
will be evaluated correctly once the LHC turns on.
\end{thebibliography}
\end{document}